\documentclass[showpacs,preprint,showkeys,superscriptaddress,amssymb,
               nofootinbib,aps]{revtex4}

\makeatletter
\renewcommand{\p@subsection}{}
\renewcommand{\p@subsubsection}{}
\makeatother
\usepackage{array,multirow}
\usepackage{amsmath}
\usepackage{amsthm}

\usepackage{amssymb}
\usepackage{bm}
\usepackage{blindtext}
\usepackage[svgnames]{xcolor}
\usepackage{enumitem}
\usepackage{graphicx}
\def\ni{\noindent}
\def\be{\begin{equation}}
\def\ee{\end{equation}}
\def\lb{\label}
\def\pr{\protect}
\DeclareMathOperator{\sech}{sech}

\begin{document}

\title{\Large Fluid dynamics in the warp drive spacetime geometry}
\author{Osvaldo L.\ Santos-Pereira}\email{olsp@if.ufrj.br}
\affiliation{Physics Institute, Universidade Federal do Rio de Janeiro, Brazil}
\author{Everton M.\ C.\ Abreu}\email{evertonabreu@ufrrj.br}
\affiliation{Physics Department, Universidade Federal Rural do Rio de
            Janeiro, Serop\'edica, Brazil}
\affiliation{Physics Department, Universidade Federal de Juiz de Fora, Brazil}
\affiliation{Graduate Program in Applied Physics, Physics Institute,
	    Universidade Federal do Rio de Janeiro, Brazil}
\author{Marcelo B. Ribeiro}\email{mbr@if.ufrj.br}
\affiliation{Physics Institute, Universidade Federal do Rio de Janeiro, Brazil}
\affiliation{Graduate Program in Applied Physics, Physics Institute,
	    Universidade Federal do Rio de Janeiro, Brazil}
\affiliation{Valongo Observatory, Universidade Federal do Rio de
            Janeiro, Brazil}
\date{\today}

\begin{abstract}
\noindent
The Alcubierre warp drive metric is a spacetime geometry featuring
a spacetime distortion, called warp bubble, where a massive particle
inside it acquires global superluminal velocities, or warp speeds.
This work presents solutions of the Einstein equations for the
Alcubierre metric having fluid matter as gravity source. The
energy-momentum tensor considered two fluid contents, the perfect
fluid and the parametrized perfect fluid (PPF), a tentative more
flexible model whose aim is to explore the possibilities of warp
drive solutions with positive matter density content.
Santos-Pereira et al.\ (2020) already showed that the Alcubierre
metric having dust as source connects this geometry to the Burgers
equation, which describes shock waves moving through an inviscid
fluid, but led the solutions back to vacuum. The same happened for
two out of four solutions subcases for the perfect fluid. Other
solutions for the perfect fluid indicate the possibility of warp
drive with positive matter density, but at the cost of a complex
solution for the warp drive regulating function. Regarding the PPF,
solutions were also obtained indicating that warp speeds could be
created with positive matter density. Weak, dominant, strong and
null energy conditions were calculated for all studied subcases,
being satisfied for the perfect fluid and creating constraints in
the PPF quantities such that positive matter density is also
possible for creating a warp bubble. Summing up all results,
energy-momentum tensors describing more complex forms of matter,
or field, distributions generate solutions for the Einstein
equations with the warp drive metric where negative matter density
might not be a strict precondition for attaining warp speeds.
\end{abstract}


\maketitle

\section{Introduction}

It is well known that in general relativity particles can travel
globally at superluminal speeds, whereas locally they cannot surpass
the light speed. The \textit{warp drive spacetime geometry} advanced
by Alcubierre \cite{Alcubierre1994} uses these physical properties
to propel material particles at superluminal speeds. It creates a
limited spacetime distortion, called \textit{warp bubble}, such that
the spacetime is contracted in front of it and expanded behind the
bubble as it moves along a geodesic. This warp drive metric is such
that a particle trapped inside this bubble would locally move at
subluminal speeds, whereas the bubble with the particle inside
acquires global superluminal velocities, or \textit{warp speeds}.
In the seminal paper, Alcubierre also concluded that the warp metric
would imply the violation of the energy conditions since it appeared
that negative energy density would be required for the creation of
the warp bubble.

Since this original work many authors have contributed in our
understanding of the theoretical details of the Alcubierre warp drive
metric and possible feasibility of matter particles acquiring warp
speeds. Ford and Roman \cite{FordRoman1996} applied quantum
inequalities to calculate the amount of negative energy required to
transport particles at superluminal speeds. They concluded that such
energy requirements would be huge, so  the amount of negative energy
density necessary for the practical construction of a warp bubble
would be impossible to achieve. Pfenning and Ford \cite{Pfenning1997}
also used quantum inequalities to calculate the necessary bubble
parameters and energy for the warp drive viability, reaching at an
enormous amount of energy, ten orders of magnitude greater than the
mass-energy of the entire visible Universe, also negative. Hiscock
{\cite{hiscock}} computed the vacuum energy-momentum tensor (EMT) of
a reduced two dimensional quantized scalar field of the warp drive
spacetime. He showed that in this reduced context that the EMT
diverges if the apparent velocity of the bubble is greater than the
speed of light. Such divergence is connected to the construction of
an horizon in this two dimensional spacetime. Due to the semiclassical
effects, the  superluminal travel via warp drive might unfeasible.
For example, to an observer within the warp drive bubble, the backward
and forward walls look  like the horizon of a white hole and of a
black hole, respectively, resulting with a Hawking radiation.

The issue of superluminal speeds of massive particles traveling
faster than photons has also been studied by Krasnikov
\cite{Krasnikov1998}, who argued that this would not be possible if
some conjectures for globally hyperbolic spacetimes are made. He
described some spacetime topologies and their respective need of the
tachyon existence for the occurrence of travel at warp speeds. This
author also advanced a peculiar spacetime where superluminal travel
would be possible without tachyons, named as the \textit{Krasnikov
tube} by Everett and Roman \cite{EveretRoman1997}, who generalized
the metric designed by Krasnikov by proposing a tube in the direction
of the particle's path providing a connection between Earth and a
distant star. Inside the tube the spacetime is flat and the lightcones
are opened out in order to allow the one direction superluminal
travel. For the Krasnikov tube to work they showed that huge quantities
of negative energy density would also be necessary. Since the tube does
not possess closed timelike curves, it would be theoretically possible
to design a two way non-overlapping tube system such that it would work
as a time machine. In addition, the EMT is positive in some regions.
Both the metric and the obtained EMT were thoroughly analyzed in Refs.\
\cite{Lobo2002,Lobo2003}.

A relevant contribution to warp drive theory was made by van de
Broeck \cite{Broeck1999}, who demonstrated that a small modification
of the original Alcubierre geometry greatly diminishes to a few solar
masses the total negative energy necessary for the creation of the
warp bubble distortion, a result that led him to hypothesize that
other geometrical modifications of this type could further reduce
in a dramatic fashion the amount of energy necessary to create a warp
drive bubble. Natario \cite{Natario2002} designed a new warp drive
concept with zero expansion by using spherical coordinates and the
$X$-axis as the polar one. Lobo and Visser \cite{LoboVisser2004b,
LoboVisser2004} discussed that the center of the warp bubble, as
proposed by Alcubierre, need to be massless (see also \cite{White2003,
White2011}). A linearized model for both approaches was introduced
and it was demonstrated that for small speeds the amassed negative
energy inside the warp field is a robust fraction of the particle's
mass inside the center of the warp bubble. Lee and Cleaver
\cite{cleaver1,cleaver2} have looked at how external radiation might
affect the Alcubierre warp bubble, possibly making it energetically
unsustainable, and how a proposed warp field interferometer could
not detect spacetime distortions. Mattingly et al.\ \cite{cleaver3,
cleaver4} discussed curvature invariants in the Natario and Alcubierre
warp drives.

In a previous paper \cite{nos} we have considered some of 
these issues, but from a different angle. Since the Alcubierre metric
was not advanced as a solution of the Einstein equations, as it was 
originally proposed simply as an ad hoc geometry designed to create a
spacetime distortion such that a massive particle inside it travels at
warp speeds, whereas locally it never exceeds the light speed, the basic
question was then the possible types of matter-energy sources capable
of creating a warp bubble. To answer this question the Einstein
equations have to be solved with some form of EMT as source. The
simplest one to start with is incoherent matter. Following this line
of investigation, we showed that the dust solutions of the Einstein
equations for the warp drive metric implied in vacuum, that is, such
distribution is incapable of creating a warp bubble, nevertheless, the
Burgers equation appeared as part of the solution of the Einstein
equations. In addition, since the Burgers equation describes shock
waves moving in an inviscid fluid, it was also found that at these
shock waves behave as plane waves \cite{Trefethen2001,Evans2010,
Forsyth1906,Bateman1915,Burgers1948}.

In this paper we generalize the results obtained in Ref.\ \cite{nos} by
following the next logical step, that is, investigating perfect fluid
as EMT source for the Alcubierre metric. We also propose a slightly
generalized perfect fluid EMT, called here \textit{parametrized perfect
fluid} (PPF), in order to produce a tentatively more flexible model
such that the pressure may have different parameter values. The aim
is to see if more flexible EMT distributions could relax the original
requirement that warp speeds could only be achieved by means of negative
matter density.

For the perfect fluid EMT solutions we found that two out of four
subcases turn out to be the dust solution of Ref.\ \cite{nos} where
both the matter density and pressure vanish, but the Burgers equation
also appears as a result of the solutions of the Einstein equations
\cite{turcos}.
Two other subcases, however, indicate that warp speeds are possible
with positive matter density, but at the cost of a complex solution for
the warp metric regulating function. Weak, dominant, strong and null
energy conditions were calculate for both EMTs and all perfect fluid
solutions satisfy them. In the case of the PPF, two out of four
solutions give rise to a nonlinear equation of state linking various
pressures to the matter density. Other solutions produced results where
a nonvanishing pressure occurs with a vanishing matter density, condition
considered unphysical and then dismissed. The solutions also produced
parameters and equations of state related to pressure and inequalities
that satisfy all the energy conditions. These results indicate that
energy-momentum tensors describing more complex forms of matter
distributions generate solutions for the Einstein equations with the
warp drive metric where negative matter density might not be a strict
precondition. 

The plan for the paper is as follows. In Section \ref{eeqs} we briefly
review the Alcubierre warp drive theory and present the relevant
equations and all nonzero components of the Einstein tensor for the
warp drive metric. In Section \ref{sec3} the Einstein equations are
then solved and solutions presented for the warp drive metric having
a perfect fluid gravity source. In Section \ref{diffluid} the non-zero
components of the Einstein tensor in the warp drive geometry are
written in terms of the PPF EMT. Solutions for this more flexible EMT
are also obtained and studied in all subcases. Section \ref{divemts}
presents the EMT divergence of both the perfect fluid and PPF, whereas
Section \ref{engconds} discusses the energy conditions for the two
types of EMTs. Section {\ref{furdisc}} provides further discussions
on the results presented in the previous sections, and Section
\ref{conc} depicts our conclusions. 


\section{Einstein Equations}\label{eeqs}
\renewcommand{\theequation}{2.\arabic{equation}}
\setcounter{equation}{0}

We shall start this section by brief reviewing the Alcubierre warp drive
metric. Subsequently, the nonzero components of the Einstein tensor of
this metric will also be explicitly shown. The expressions presented in
this section form the basic set of equations required for the next
sections. 

\subsection{The Alcubierre warp drive geometry}

The start up geometry advanced in Ref.\ \cite{Alcubierre1994} may be
written as follows,
\be
{ds}^2 = - \left(\alpha^2 -\beta_i\beta^{i}\right) \, dt^2 
+ 2 \beta_i \, dx^i \, dt + \gamma_{ij} \, dx^i \, dx^j \,\,,
\label{metric1}
\ee
where $d\tau$ is the  proper time lapse, $\alpha$ is the lapse 
function, $\beta^i$ is the spacelike shift vector and $\gamma_{ij}$ 
is the spatial metric for the hypersurfaces.\footnote{Throughout this
paper Greek indices will range from 0 to 3, whereas the Latin ones
indicate the spacelike hypersurfaces and will range from 1 to 3.} The
lapse function $\alpha$ and the shift vector $\beta^i$ are functions
to be determined, whereas $\gamma_{ij}$ is a positive-definite metric
on each of the spacelike hypersurfaces, for all values of time, a
feature that makes the spacetime globally hyperbolic \cite{Alcubierre2012,
DeWitt1979}.

Alcubierre \cite{Alcubierre1994} assumed the following particular
parameter choices for Eq.\,\eqref{metric1},
\begin{align}
\alpha &= 1, 
\\
\beta^1& = - v_s(t)f\big[r_s(t)\big], \label{betax}
\\
\beta^2 &= \beta^3 = 0,
\\
\gamma_{ij} &= \delta_{ij}.
\end{align}
Hence, the \textit{Alcubierre  warp drive metric} is given by,
\begin{equation}
ds^2 = -\left[1 - v_s(t)^2 f(r_s)^2\right]dt^2-v_s(t) f(r_s)\,dx\,dt 
+ dx^2 + dy^2 + dz^2 \,\,,
\label{alcmetric1}
\end{equation}
where $v_s(t)$ is the velocity of the center of the bubble moving
along the curve $x_s(t)$. This is given by the following expression,
\begin{equation}
v_s(t) = \frac{dx_s (t)}{dt}\,\,.
\end{equation} 
The function $f(r_s)$ is the warp drive \textit{regulating function}.
It describes the shape of the warp bubble, which is given by the 
following expression \cite{Alcubierre1994},
\begin{equation}
f(r_s) = \frac{\tanh\left[\sigma(r_s + R)\right] 
- \tanh\left[\sigma(r_s - R)\right]}
{2 \tanh(\sigma R)} \,\,,
\label{regfunction}
\end{equation}
where $\sigma$ and $R$ are parameters to be determined. The variable 
$r_s(t)$ defines the distance from the center of the bubble
$[x_s(t),0,0]$ to a generic point $(x,y,z)$ on the surface of the 
bubble, given by the following equation,
\begin{equation}
r_s(t) = \sqrt{\left[x - x_s(t)\right]^2 + y^2 + z^2}.
\lb{rsxyz}
\end{equation} 
From the above one can see that the motion is one-dimensional, since
the $x$-coordinate is the only one perturbed by the function $x_s(t)$.

\subsection{Einstein tensor components}

Let us now adopt Alcubierre's original notation by assuming
\be
\beta = -\beta^1=v_s(t)f(r_s)
\lb{betax2}
\ee
in Eq.\ \eqref{betax}. Then, the nonzero components of the Einstein
tensor for the metric (\ref{alcmetric1}) are given by the following
expressions:
\begin{eqnarray}
G_{00} &=&  - \frac{1}{4} 
(1 + 3\beta^2)
\left[
\left(\frac{\partial \beta}{\partial y} \right)^2 +  
\left(\frac{\partial \beta}{\partial z} \right)^2 
\right] 
- \beta \left(\frac{\partial^2 \beta}{\partial y^2} + 
\frac{\partial^2 \beta}{\partial z^2}\right),
\label{et00}
\\[2pt]
G_{01} &=&  \frac{3}{4} 
\beta \left[
\left(\frac{\partial \beta}{\partial y}\right)^2 
+ \left(\frac{\partial \beta}{\partial z}\right)^2 
\right] 
+ \frac{1}{2}\left(
\frac{\partial^2 \beta}{\partial y^2} 
+ \frac{\partial^2 \beta}{\partial z^2}
\right),
\label{et01}
\\[4pt]
G_{02} &=& - \frac{1}{2}
\frac{\partial^2 \beta}{\partial x \partial y} 
- \frac{\beta}{2} 
\left(2\frac{\partial \beta}{\partial y}
\, \frac{\partial \beta}{\partial x} +
\beta \frac{\partial^2 \beta}{\partial x \partial y} +
\frac{\partial^2 \beta}{\partial t \partial y}\right),
\label{et02}
\\[6pt]
G_{03} &=& - \frac{1}{2}
\frac{\partial^2 \beta}{\partial x \partial z} 
- \frac{\beta}{2} 
\left(2\frac{\partial \beta}{\partial z}
\, \frac{\partial \beta}{\partial x} +
\beta \frac{\partial^2 \beta}{\partial x \partial z} +
\frac{\partial^2 \beta}{\partial t \partial z}\right),
\label{et03}
\end{eqnarray}
\begin{eqnarray}
%
G_{11} &=& - \frac{3}{4} \left[
\left(\frac{\partial \beta}{\partial y}\right)^2 
+ \left(\frac{\partial \beta}{\partial z}\right)^2
\right], \label{et11}
\\[2pt]
G_{12} &=& \frac{1}{2}\left(
2 \frac{\partial \beta}{\partial y} \, 
\frac{\partial \beta}{\partial x} 
+ \beta \frac{\partial^2 \beta}{\partial x \partial y} 
+ \frac{\partial^2 \beta}{\partial t \partial y}\right),
\label{et12}
\\[2pt]
G_{13} &=& \frac{1}{2}\left(
2 \frac{\partial \beta}{\partial z} \, 
\frac{\partial \beta}{\partial x} 
+ \beta \frac{\partial^2 \beta}{\partial x \partial z} 
+ \frac{\partial^2 \beta}{\partial t \partial z}\right),
\label{et13}
\\[2pt]
G_{23} &=& \frac{1}{2} \frac{\partial \beta}{\partial z} 
\, \frac{\partial \beta}{\partial y},
\label{et23}
\\[2pt]
G_{22} &=& - \left[
\frac{\partial^2 \beta}{\partial t \partial x}
+ \beta \frac{\partial^2 \beta}{\partial x^2}
+ \left(\frac{\partial \beta}{\partial x}\right)^2
\right]
- \frac{1}{4}\left[
\left(\frac{\partial \beta}{\partial y}\right)^2
- \left(\frac{\partial \beta}{\partial z}\right)^2
\right],
\label{et22}
\\[2pt]
G_{33} &=& - \left[
\frac{\partial^2 \beta}{\partial t \partial x}
+ \beta \frac{\partial^2 \beta}{\partial x^2}
+ \left(\frac{\partial \beta}{\partial x}\right)^2
\right]
+ \frac{1}{4}\left[
\left(\frac{\partial \beta}{\partial y}\right)^2
- \left(\frac{\partial \beta}{\partial z}\right)^2
\right]\,\,.
\label{et33}
\end{eqnarray}


\section{Perfect fluid}\label{sec3}
\renewcommand{\theequation}{3.\arabic{equation}}
\setcounter{equation}{0}

Besides incoherent matter, or dust, already studied in Ref.\ \cite{nos},
the simplest matter-energy distribution to be considered as gravity
source for the possible creation of a warp bubble, and then warp speeds,
is the perfect fluid. Hence, this section will discuss matter content
solutions of the Einstein equations considering a perfect fluid matter
source EMT for the Alcubierre metric.

\subsection{Perfect fluid content solutions}

The EMT for a perfect fluid may be written as follows,
\begin{equation}
T_{\alpha \beta} = \left(\mu + p\right) \, u_\alpha u_\beta
+ p \, g_{\alpha \beta},
\end{equation}
where $\mu$ is the matter density, $p$ is the fluid pressure, $g_{\alpha
\beta}$ is the metric tensor and $u_{\alpha}$ is the 4-velocity of an
observer inside the fluid. Perfect fluids have no shear stress,
rotation, heat conduction or viscosity, nevertheless this ideal fluid
provides a more complex matter content than simple dust \cite{Schutz2009},
allowing us to study if a warp bubble can be created with this gravity
source and how the respective gravity field equations solutions can be
understood.

For the metric (\ref{alcmetric1}) the perfect fluid EMT assumes the
following form,
\begin{equation}
T_{\alpha\beta} = 
\begin{pmatrix} 
\mu + \beta^2 p & - \beta p   & 0 & 0  \\ 
- \beta p         & p         & 0 & 0  \\ 
0                 & 0         & p & 0  \\ 
0                 & 0         & 0 & p 
\end{pmatrix}\,.
\label{pfemt}
\end{equation}
Let us now use Eqs.\ \eqref{et00} to \eqref{et33} with the EMT above
in the Einstein equations. Substituting components $G_{11} = 8\pi
T_{11}$ and $G_{01} = 8\pi T_{01}$ into $G_{00} = 8\pi T_{00}$, after
some  algebra and simplifications we may write the following expression,
\begin{equation}
T_{00} + 2 \beta T_{01} + \frac{1}{3}(3\beta^2 - 1)T_{11} = 0\,\,.
\label{eqstaterel}
\end{equation}
Substituting the values for the EMT components $T_{00} = \mu+\beta^2 p$,
$T_{01} = - \beta p$, $T_{11} = p$ Eq.\ \eqref{eqstaterel} results in
the following expression,
\begin{equation}
p = 3 \mu.
\label{eqstateeq}
\end{equation}
This is an equation of state for the Alcubierre metric having a
perfect fluid gravity source EMT.

The component $G_{23}$ is zero since $T_{23} = 0$. This case
leads us to either $\partial \beta/\partial y$, or $\partial \beta/ 
\partial z$, or both, equal to zero. Let us now analyze these 
possibilities and its consequences.

\begin{description}[align=left]
\item[Case 1: \small $\bm{\left[\displaystyle\frac{\partial\beta}
	{\partial z}=0\right]}$]
As $\beta $ does not depend on $z$, the Einstein tensor components
$G_{13}$, $G_{23}$ and $G_{03}$ are identically zero. Substituting
this case into $G_{11}=8\pi T_{11}$, where $G_{11}$ is given by Eq.\
\eqref{et11} and $T_{11} = p$, it follows immediately the result
below,
\begin{equation}
\frac{3}{4}\left(\frac{\partial \beta}{\partial y}\right)^2 =
	- 8 \pi p.
\end{equation}
Substituting $(\partial \beta/\partial y)^2$ above into $G_{01} =
8 \pi T_{01}$, where $G_{01}$ is given by Eq.\ \eqref{et01} and
$T_{01} = - \beta p$ from Eq.\ \eqref{pfemt}, as well as $G_{12}
= 8 \pi T_{12}$ and $G_{02} = 8 \pi T_{02}$, where $T_{12} = 0$
and $T_{02} = 0$, the Einstein equations are reduced to the
following equations,
\begin{eqnarray} 
&& p = 3 \mu,
\label{eepf01}
\\ [2pt]
&& \left(\frac{\partial \beta}{\partial y}\right)^2 
= - \frac{32}{3} \pi p =-\,32\pi\,\mu,
\label{eepf02}
\\ [2pt]
&& \frac{\partial^2 \beta}{\partial y^2} = 0,
\label{eepf03}
\\ [2pt]
&& \frac{\partial^2 \beta}{\partial x\partial y}=0,
\label{eepf04}
\\ [2pt]
&& 2 \frac{\partial \beta}{\partial y} \, 
\frac{\partial \beta}{\partial x} 
+ \frac{\partial^2 \beta}{\partial t \partial y}=0,
\label{eepf05}
\\ [2pt]
&& \frac{\partial^2 \beta}{\partial t \partial x}
+ \beta \frac{\partial^2 \beta}{\partial x^2}
+ \left(\frac{\partial \beta}{\partial x}\right)^2
= - \frac{64}{3} \pi p \,=\, - \,64\pi\mu,
\label{eepf06}
\\ [2pt]
&& \frac{\partial^2 \beta}{\partial t \partial x}
+ \beta \frac{\partial^2 \beta}{\partial x^2}
+ \left(\frac{\partial \beta}{\partial x}\right)^2
= - \frac{128}{3} \pi p\,=\,-\,128\pi\mu\,\,.
\label{eepf07}
\end{eqnarray}

Eq.\ \eqref{eepf02} implies that $\partial \beta / \partial y$ must be
constant, since the pressure $p$ is assumed constant. Eq.\ \eqref{eepf03}
also shows that $\beta$ must be a linear function of the $y$-coordinate,
which means that $\beta$ must have a possible additional dependence of
arbitrary functions on $t$ and $x$. Both expressions in Eqs.\
\eqref{eepf06} and \eqref{eepf07} constitute the same homogeneous partial
differential equation, but with different inhomogeneous parts, so the
solution of the inhomogeneous equation is not unique, unless the pressure
$p$ is zero. Then, considering these points and Eqs.\ \eqref{eepf02} and
\eqref{eepf05}, it follows that,
\begin{equation}
\frac{\partial \beta}{\partial y} \, 
\frac{\partial \beta}{\partial x} = 0,
\label{subs1a}
\end{equation}
which means that either of these partial derivatives, or both, vanish.
Let us discuss both possibilities below. 

\item[Case 1a: \small $\bm{\left[\displaystyle \frac{\partial
     \beta}{\partial z} =0 \,\,\,\, \text{and} \,\,\,\, \frac{\partial
     \beta}{\partial x} =0\right]}$]

For this case the set of partial differential equations from Eqs.\
\eqref{eepf01} to \eqref{eepf07} simplify to,
\begin{align} 
&p = 3 \mu,
\\
&\frac{\partial \beta}{\partial y}
=\pm\sqrt{- 32 \pi \mu}.
\end{align}
The above equations mean that the matter density $\mu$ must be negative
or zero for a non complex solution of the Einstein equations, and
$\beta$ must be a function of both $t$ and $x$ coordinates only. The
equation above is readily integrated, yielding 
\begin{equation} 
\beta(t,y) =\pm\sqrt{- 32 \pi \mu} \, y + g(t)\,\,,
\label{beta1a}
\end{equation}
where $g(t)$ is a function to be determined by the boundary 
conditions. 

\item[Case 1b: \small $\bm{\left[\displaystyle \frac{\partial
     \beta}{\partial z} =0 \,\,\,\, \text{and} \,\,\,\, \frac{\partial
     \beta}{\partial y} =0\right]}$]
     
In this case the pressures vanishes, since $\partial \beta/\partial
y =0$, and the set of partial differential equations \eqref{eepf01}
to \eqref{eepf07} simplify to,
\begin{align} 
&p = 3 \mu=0,
\\
&\frac{\partial^2 \beta}{\partial t \partial x}
+ \beta \frac{\partial^2 \beta}{\partial x^2}
+ \left(\frac{\partial \beta}{\partial x}\right)^2
= 0.
\label{neweepf07}
\end{align}
Therefore, for $p=0$ the equation of state $p = 3 \mu$ implies zero
matter density as well, which reduces the solution to the dust case
and then vacuum. This also leads to the appearance of shock waves
as plane waves, since $\beta=\beta(x,t)$ and the field equations are
reduced to the Burgers equation, as studied in Ref.\ \cite{nos}.
This is the case because Eq.\ \eqref{neweepf07} can be written in the
following form,
\begin{equation}
\frac{\partial \beta}{\partial t}
+ \frac{1}{2}\frac{\partial}{\partial x} (\beta^2)
= h(t).   
\label{pfadvec4}
\end{equation}
Here $h(t)$ is a generic function to be determined by boundary
conditions. In its homogeneous form, where $h(t) = 0$, it takes the
conservation form of the inviscid Burgers equation.
\be
\frac{\partial \beta}{\partial t}
+ \frac{1}{2}\frac{\partial}{\partial x} (\beta^2)
= 0\,\,.   
\label{invburgeq}
\ee
See Ref.\ \cite{nos} for details of the Burgers equation in this context.

\item[Case 2: \small $\bm{\left[\displaystyle\frac{\partial\beta}
	{\partial y}=0\right]}$]

As $\beta $ does not depend on the $y$-coordinate, it is easy to see
that $G_{12}$, $G_{23}$ and $G_{02}$ are identically zero. In addition,
considering this case into $G_{11} = 8 \pi T_{11}$, it follows
immediately that,
\begin{equation}
- \frac{3}{4} \left(\frac{\partial \beta}{\partial z}\right)^2=8\pi p.
\label{aaa1}
\end{equation}
Substituting Eq.\ (\ref{aaa1}) in $G_{01}=8\pi T_{01}$, where $G_{01}$ is 
given by Eq.\ \eqref{et01}, $T_{01}=- \beta p$ from Eq.\ \eqref{pfemt},
and inserting the component $G_{13}=8 \pi T_{13}$ into the component
$G_{03} = 8 \pi T_{03}$, where $T_{13}=T_{03} = 0$, after some algebra
we reach at the following expressions,
\begin{align} 
&p = 3 \mu,
\label{eepf01c2}
\\
&\left(\frac{\partial \beta}{\partial z}\right)^2 
= - \frac{32}{3} \pi p,
\label{eepf02c2}
\\ 
&\frac{\partial^2 \beta}{\partial z^2} = 0,
\label{eepf03c2}
\\
&\frac{\partial^2 \beta}{\partial x\partial z} = 0,
\label{eepf04c2}
\\
&2 \frac{\partial \beta}{\partial z} \, 
\frac{\partial \beta}{\partial x} 
+ \frac{\partial^2 \beta}{\partial t \partial z} = 0,
\label{eepf05c2}
\\
&\frac{\partial^2 \beta}{\partial t \partial x}
+ \beta \frac{\partial^2 \beta}{\partial x^2}
+ \left(\frac{\partial \beta}{\partial x}\right)^2
= - \frac{64}{3} \pi p,
\label{eepf06c2}
\\
&\frac{\partial^2 \beta}{\partial t \partial x}
+ \beta \frac{\partial^2 \beta}{\partial x^2}
+ \left(\frac{\partial \beta}{\partial x}\right)^2
= - \frac{128}{3} \pi p.
\label{eepf07c2}
\end{align}

Eq.\ \eqref{eepf03c2} shows that $\beta$ is a linear function with
respect to the $z$-coordinate. Eq.\,\eqref{eepf02c2} implies that
$\partial \beta / \partial z$ must be constant since the pressure
$p$ is assumed to be a constant. This means that all second partial
derivatives of $\partial \beta / \partial z$ must vanish. Eqs.\
\eqref{eepf06c2} and \eqref{eepf07c2} are the same homogeneous
partial differential equation, but both right hand side of theirs
are different. Hence, the solution of the non-homogeneous equation
is not unique, unless the pressure $p$ is zero. Considering Eq.\
\eqref{eepf05c2}, this case also unfolds in two possibilities,
since,
\begin{equation}
\frac{\partial \beta}{\partial z} \, 
\frac{\partial \beta}{\partial x} = 0,
\end{equation}
and either or both are zero. Let us now analyze each subcase.

\item[Case 2a: $\bm{\displaystyle \left[\frac{\partial\beta}{\partial 
y} = 0 \ \text{and} \ \frac{\partial \beta}{\partial x} = 0\right]}$]

For this case Eqs.\ \eqref{eepf01c2} to \eqref{eepf07c2} yield,
\begin{align} 
&p = 3 \mu\,,
\label{p3mu}
\\
&\frac{\partial \beta}{\partial z}
= \pm \,\sqrt{- 32 \pi \mu}\,, 
\label{bbb1}
\\
&\frac{\partial \beta}{\partial z}
= \pm \,\sqrt{\pm 96 \pi \mu}\,.
\label{bbb2}
\end{align}

Eq.\ (\ref{bbb1}) means that the matter density $\mu$ must be negative
or zero for a non complex solution of the Einstein equations. Eq.\
(\ref{bbb2}) allows for possible positive matter density. In addition,
$\beta$ has its dependence reduced to $\beta=\beta(z,t)$. If the matter
density $\mu$ is assumed constant, then the above expressions can be
integrated, yielding 
\begin{align} 
&\beta(z,t) = \pm\sqrt{- 32 \pi \mu} \, z + \bar{g}(t)\,,
\label{beta2a}
\\
&\beta(z,t) = \pm\sqrt{\pm 96 \pi \mu} \, z + \bar{h}(t)\,.
\label{beta2a2}
\end{align}
where $\bar{g}(t)$ and $\bar{h}(t)$ are arbitrary functions to be
determined by boundary conditions. 

\item[Case 2b $\bm{\displaystyle \left[\frac{\partial \beta}{\partial 
y} = 0 \ \text{and} \ \frac{\partial \beta}{\partial z} = 0\right]}$]

For this subcase Eq.\ (\ref{eepf02c2}) implies zero pressure, and the
set of partial differential equations \eqref{eepf01c2} to
\eqref{eepf07c2} simplify to,
\begin{align} 
&p = 3 \mu=0,
\\
&\frac{\partial^2 \beta}{\partial t \partial x}
+ \beta \frac{\partial^2 \beta}{\partial x^2}
+ \left(\frac{\partial \beta}{\partial x}\right)^2
= 0\,\,,
\label{neweepf07c2}
\end{align}
where the last equation is the result of the only nonzero Einstein tensor
components $G_{22}$ and $G_{33}$. These results are the same as Case 1b
above, that is, the dust solution for the Alcubierre warp drive metric
that results in the Burgers equation (\ref{pfadvec4}) and its inviscid
form given by Eq.\ (\ref{invburgeq}), as well as shock waves as plane
waves \cite{nos}.
\end{description}
\vspace{-0.6cm} $\square$ \vspace{0.3cm}

Table \ref{tab1} summarizes all cases and their respective results of 
the Einstein equations with the Alcubierre warp drive metric having a
perfect fluid matter content as gravity source.

\begin{table}
\begin{tabular}{| m{3cm} | m{3cm} | m{7cm} |}
\hline 
Case & Condition & Results \\ 
\hline 
\multirow{2}{*}{$1) \
\displaystyle{\frac{\partial \beta}{\partial z} = 0}$}
& 
$1a) \ \displaystyle{\frac{\partial \beta}{\partial x} = 0}$
&
$\begin{array} {ll} 
p = 3 \mu \\ [6pt]
\beta = \beta(y,t)\\ [6pt]
\displaystyle{\frac{\partial \beta}{\partial y} 
=\pm\sqrt{-32 \pi \mu}} \\ [8pt]
\end{array}$ \\ [28pt]
\cline{2-3}   
&
$1b) \ \displaystyle{\frac{\partial \beta}{\partial y} = 0}$
&
$\begin{array} {ll} 
p = 3 \mu = 0 \\ [6pt]
\beta = \beta(x,t)\\ [6pt]
\displaystyle{\frac{\partial \beta}{\partial t} 
+ \frac{1}{2} \frac{\partial}{\partial x}(\beta^2)
= h(t)} \\ [8pt]
\rightarrow\mbox{this is the dust solution of Ref.\ \cite{nos}} \\ [4pt]
\end{array}$ \\ [28pt] 
\hline 

\multirow{2}{*}{$2) \
\displaystyle{\frac{\partial \beta}{\partial y} = 0}$}
& 
$2a) \ \displaystyle{\frac{\partial \beta}{\partial x} = 0}$
&
$\begin{array} {ll} 
p = 3 \mu \\ [6pt]
\beta = \beta(z,t)\\ [6pt]
\displaystyle{\frac{\partial \beta}{\partial z} 
= \pm\,\sqrt{-32 \pi \mu}}\\ [8pt]
\displaystyle{\frac{\partial \beta}{\partial z} 
=\,\pm \sqrt{\pm\,96 \pi \mu}} \\ [8pt]
\end{array}$ \\ [28pt]
\cline{2-3}   
&
$2b) \ \displaystyle{\frac{\partial \beta}{\partial z} = 0}$
&
$\begin{array} {ll} 
p = 3 \mu = 0 \\ [6pt]
\beta = \beta(x,t)\\ [6pt]
\displaystyle{\frac{\partial \beta}{\partial t} 
+ \frac{1}{2} \frac{\partial}{\partial x}(\beta^2)
= h(t)} \\ [6pt]
\rightarrow\mbox{this is the dust solution of Ref.\ \cite{nos}} \\ [4pt]
\end{array}$ \\ [28pt] 
\hline 
\end{tabular}
\caption{Summary of all solutions of the Einstein equations with
         the Alcubierre warp drive metric having perfect fluid EMT
         as mass-energy source.}
\label{tab1}
\end{table}

\subsection{Discussion}\label{discu}

Cases 1b and 2b are simply the dust case already studied in Ref.\
\cite{nos}, apparently being unable to generate a warp bubble since
this is a vacuum solution, although it connects the warp metric to
the Burgers equation and then to shock waves in the form of plane
waves.

Cases 1a and 2a share the same equation of state $p = 3 \mu$, but
the coordinate dependencies are different, since $\beta=\beta(y,t)$
and $\beta=\beta(z,t)$, respectively. For $\beta$ to be a real valued
function the matter density $\mu$ must be negative in Case 1a. From
Eqs.\ \eqref{beta1a} and \eqref{beta2a} we have assumed a constant
matter density, which means a straightforward integration, but $\mu$
can be also a function of both $t$ and $y$ coordinates in the Case 1a,
or a function of both $t$ and $z$ coordinates in the Case 2a.

Inasmuch as the matter density must be negative for real solutions,
one could think of defining the total mass-energy density as follows,
\begin{align}
&\mu(t,x^j) = \mu^+ + a(t,x^j) \mu^- \leq 0, 
\label{mattnegpos}\\
&\nonumber x^j=y,z,
\end{align}
where $\mu^+$ is the positive portion of the matter density of the
perfect fluid and $\mu^-$ its negative portion that would allow the
warp bubble to exist. $a(t,x^j)$ would be a regulating function that
depends on both time $t$ and space $x^j$ coordinates, being related
to the shape and location of the bubble. Remembering that $x^j = y$
for the Case 1a and $x^j = z$ for the Case 2a, since $\mu(t,x^j)$
must be negative there would be a restriction for the positive and
negative portions of the matter density in Eq.\ \eqref{mattnegpos}.

It might be argued that there is no problem in complex solutions for
$\beta$, but in the warp drive scenario $\beta = v_s(t) f(r_s)$
determines both the velocity and the shape of the bubble, so either
the velocity $v_s(t)$ or the regulating function $f(r_s)$ of the
bubble shape must be complex. A complex velocity has no physical
meaning, but a complex regulating function could be acceptable if
we only consider its real part. Thus, in such situation the formation
of a warp bubble capable of generating warp speeds could still be
possible in the presence of a perfect fluid positive matter density
EMT as gravity source.

Nevertheless, caution is required here because if the result of
integrating $\beta$ turns out to be purely imaginary it is not clear
what the bubble shape being represented by an imaginary function
means. Therefore, in principle it seems reasonable to start with
$\beta$ being a real function and the warp bubble requiring a perfect
fluid with negative mass-energy density for warp speeds to be
physically viable. But, it seems to us that this point remains open
to debate.

Considering the results above it is apparent that a perfect fluid
EMT generated a more complex set of solutions of the Einstein
equations than the dust one, then it is conceivable that viable
warp speeds are also possible in more complex EMTs. One such
possibility will be discussed next.


\section{Parametrized perfect fluid} \label{diffluid}
\renewcommand{\theequation}{4.\arabic{equation}}
\setcounter{equation}{0}
	 
Let us propose a generalization of the perfect fluid EMT having seven
quantities, namely the mass-energy density $\mu$, the $\beta$ function
and five different pressures $A, B, C, D$ and $p$. The last quantity $D$
is a momentum density parameter. In the perfect fluid of Eq.\
\eqref{pfemt} the pressure denoted by $p$ are all the same in the EMT,
a constraint that has been relaxed here. Let us call the perfect fluid
generalization with the quantities above as the \textit{parametrized
perfect fluid} (PPF). Its respective EMT may be written as below,
\begin{equation}
T_{\alpha \sigma} = 
\begin{pmatrix} 
\mu + \beta^2 p & - \beta D & 0  & 0  \\ 
- \beta D     & A         & 0  & 0  \\ 
0             & 0         & B  & 0  \\ 
0             & 0         & 0  & C 
\end{pmatrix}\,.
\label{71}
\end{equation}
The quantities $A, B, C, D$ and $p$ will not be assumed as constants,
but rather as functions of the spacetime coordinates $(t,x,y,z)$.

This is clearly a more flexible EMT than the perfect fluid, and it
is being proposed here as a tentative model in order to explore the
consequences of more complex EMTs in terms of generating possible
positive matter solutions of the Einstein equations with the warp
drive metric without the caveats of the perfect fluid solutions
discussed above. It is a tentative proposal for a more flexible,
or toy, model for the possible creation of warp bubbles, and then
warp speeds. Section \ref{anifluid} provides more details on the
physics of this specific fluid proposal.

The nonzero components of the Einstein equations for the PPF are given
by the following expressions, 
\begin{eqnarray}
-\frac{1}{4}(1+3\beta^2)\left[ \left(\frac{\partial \beta}{\partial y}
	\right)^2 + \left(\frac{\partial \beta}{\partial z} \right)^2
        \right]-\beta \left(\frac{\partial^2 \beta}{\partial y^2}+
	\frac{\partial^2 \beta}{\partial z^2}\right) &=& 8 \pi (\mu +
	\beta^2 p),
\label{G00}
\\ [2pt]
\frac{3}{4} \beta \left[ \left(\frac{\partial \beta}{\partial y}\right)^2
	+\left(\frac{\partial \beta}{\partial z}\right)^2 \right] +
	\frac{1}{2}\left( \frac{\partial^2 \beta}{\partial y^2} +
	\frac{\partial^2 \beta}{\partial z^2} \right) &=& - 8\pi\beta D,
\label{G01}
\\ [2pt]
-\frac{1}{2}\frac{\partial^2 \beta}{\partial x \partial y}-\frac{\beta}{2}
        \left(2\frac{\partial \beta}{\partial y} \, \frac{\partial \beta}
	{\partial x}+\beta \frac{\partial^2 \beta}{\partial x \partial y}
        +\frac{\partial^2 \beta}{\partial t \partial y}\right) &=& 0,
\label{G02}
\\ [2pt]
-\frac{1}{2}\frac{\partial^2 \beta}{\partial x \partial z}-\frac{\beta}{2}
        \left(2\frac{\partial \beta}{\partial z} \, \frac{\partial \beta}
	{\partial x}+\beta \frac{\partial^2 \beta}{\partial x \partial z}
        + \frac{\partial^2 \beta}{\partial t \partial z}\right) &=& 0,
\label{G03}
\\ [2pt]
- \frac{3}{4} \left[ \left(\frac{\partial \beta}{\partial y}\right)^2
       + \left(\frac{\partial \beta}{\partial z}\right)^2 \right] &=& 8 \pi A,
\label{G11}
\end{eqnarray}
\begin{eqnarray}
%
\frac{1}{2}\left( 2 \frac{\partial \beta}{\partial y} \, \frac{\partial
	\beta}{\partial x}+\beta \frac{\partial^2 \beta}{\partial x
	\partial y} + \frac{\partial^2 \beta}{\partial t \partial y}
\right) &=& 0,
\label{G12}
\\ [2pt]
\frac{1}{2}\left( 2 \frac{\partial \beta}{\partial z} \, \frac{\partial
	\beta}{\partial x} + \beta \frac{\partial^2 \beta}{\partial x
	\partial z} + \frac{\partial^2 \beta}{\partial t \partial z}\right)
	&=& 0,
\label{G13}
\\ [2pt]
\frac{1}{2} \frac{\partial \beta}{\partial z} \, \frac{\partial \beta}
{\partial y} &=& 0,
\label{G23}
\\ [2pt]
-\left[ \frac{\partial^2 \beta}{\partial t \partial x} +
	\beta \frac{\partial^2 \beta}{\partial x^2} + \left(\frac{\partial
	\beta}{\partial x}\right)^2 \right] - \frac{1}{4}\left[ \left(
	\frac{\partial \beta}{\partial y}\right)^2 - \left(\frac{\partial
\beta}{\partial z}\right)^2 \right] &=& 8 \pi B, 
\label{G22}
\\ [2pt]
-\left[ \frac{\partial^2 \beta}{\partial t \partial x}+\beta
	\frac{\partial^2 \beta}{\partial x^2} + \left(\frac{\partial
	\beta}{\partial x}\right)^2 \right] + \frac{1}{4}\left[ \left(
	\frac{\partial \beta}{\partial y}\right)^2 - \left(\frac{\partial
\beta}{\partial z}\right)^2 \right] &=& 8 \pi C. 
\label{G33}
\end{eqnarray}

Substituting Eqs.\ \eqref{G11} and \eqref{G01} into Eq.\ \eqref{G00} it
follows immediately that,
\begin{equation}
\mu = \beta^2 (2D - A - p) + \frac{A}{3} \,\,.    
\label{matterdens}
\end{equation}
This expression shows that the fluid density not only depends on the
pressure components $A$ and $p$, but also on the momentum component
$D$ and the warp bubble, since it varies with the shift vector $\beta$
and, hence, the bubble movement. So, the bubble shape modifies the
fluid density in a local way, a result that may imply an analogy with
classical fluid dynamics and shock waves in fluids with global velocity
greater than the speed of sound in that medium.

Applying this analogy to the warp drive, it may mean that the warp
bubble plays the role of a shock wave in a fluid that moves with
apparent velocity greater than the speed of light for an outside
observer far away from the bubble, this being a result of the
nonlinearity of the Einstein equations. The bubble modifies the fluid
density which then causes the bubble motion. This classical relativistic
fluid analogy may be a physical argument for a mechanism which accounts
for the great amount of energy necessary for the feasibility of the warp
drive.

After some algebra on the set of Einstein equations above they can be
rewritten as below, 
\begin{eqnarray} 
\beta^2 (2D - A - p) + \frac{A}{3} &=& \mu,
\label{set01}
\\ [2pt]
\frac{\partial^2 \beta}{\partial x \partial y} &=& 0,
\label{set02}
\\ [2pt]
\frac{\partial^2 \beta}{\partial x \partial z}  &=& 0,
\label{set03}
\\ [2pt]
\left(\frac{\partial \beta}{\partial y}\right)^2 + \left(\frac{\partial
\beta}{\partial z}\right)^2 &=& - \frac{32}{3} \pi A,
\label{set04}
\\ [2pt]
\left(\frac{\partial \beta}{\partial y}\right)^2 - \left(\frac{\partial
\beta}{\partial z}\right)^2 &=& 16 \pi (C-B),
\label{set05}
\\ [2pt]
2 \frac{\partial \beta}{\partial y} \, \frac{\partial \beta}{\partial x}  
+ \frac{\partial^2 \beta}{\partial t \partial y} &=& 0,
\label{set06}
\end{eqnarray}
\begin{eqnarray}
%
2 \frac{\partial \beta}{\partial z} \, \frac{\partial \beta}{\partial x} 
+ \frac{\partial^2 \beta}{\partial t \partial z} &=& 0,
\label{set07}
\\ [2pt]
\frac{\partial \beta}{\partial z} \, \frac{\partial \beta}{\partial y}
&=& 0,
\label{set08}
\\ [2pt]
\frac{\partial}{\partial x}\left[ \frac{\partial \beta}{\partial t}
+ \frac{1}{2}\frac{\partial}{\partial x}(\beta^2) \right] &=& -32 \pi
(B + C),
\label{set09}
\\ [2pt]
\frac{\partial^2 \beta}{\partial y^2} + \frac{\partial^2 \beta}{\partial
z^2} &=& 16 \pi \beta (A-D).
\label{set10}
\end{eqnarray}
Eq.\ \eqref{set08} shows that the solutions for the set of differential
equations above have similar alternative cases as in the perfect fluid
solutions, that is, either $\partial \beta/\partial z = 0$ and/or
$\partial \beta/\partial y = 0$. Both situations and their respective
subcases will be discussed next.

\begin{description}[align=left]
\item[Case 1: \small $\bm{\left[\displaystyle\frac{\partial\beta}
     {\partial z}=0\right]}$]
The set of equations \eqref{set01} to \eqref{set10} simplify to,
\begin{eqnarray} 
&& \mu = \beta^2 (2D - A - p) + \frac{A}{3},
\label{c1set01}
\\ [2pt]
&& \frac{\partial^2 \beta}{\partial x \partial y} = 0,
\label{c1set02}
\\ [2pt]
&&\left(\frac{\partial \beta}{\partial y}\right)^2 
= - \frac{32}{3} \pi A = 16 \pi (C-B),
\label{c1set04}
\\ [2pt]
&& 2 \frac{\partial \beta}{\partial y} \, \frac{\partial \beta}
{\partial x}+\frac{\partial^2 \beta}{\partial t \partial y}=0,
\label{c1set06}
\\ [2pt]
&& \frac{\partial}{\partial x}\left[ \frac{\partial \beta}{\partial t}
+ \frac{1}{2}\frac{\partial}{\partial x}(\beta^2) \right] = -32 \pi (B + C). 
\label{c1set09}
\\ [2pt]
&& \frac{\partial^2 \beta}{\partial y^2} = 16 \pi \beta (A-D).
\label{c1set10}
\end{eqnarray}

From Eq.\,\eqref{c1set04} a relation between the pressures $A, B$
and $C$ are straightforward,
\begin{equation}
B = C + \frac{2}{3}A.
\label{AAAA-B}
\end{equation}
From Eqs.\ \eqref{c1set02} and \eqref{c1set04} it is easy to verify
that $A$ and $C-B$ do not depend on the $x$-coordinate. In addition, for
real solutions $A$ must be negative, assuming only real values that
are equal to or smaller than zero. Differentiating Eq.\ \eqref{c1set04}
with respect to $x$ yields, 
\begin{equation}
2 \frac{\partial \beta}{\partial y} \frac{\partial^2 \beta}{\partial
x \partial y} = 0\,
\Longrightarrow \frac{\partial \beta}{\partial y} = 0 \quad
\mbox{and/or} \quad \frac{\partial^2 \beta}{\partial x \partial y} = 0,
\label{AAAA-A}
\end{equation}
and inserting the result $\partial \beta / \partial y=0$  into Eq.\
\eqref{c1set06} it follows that,
\begin{equation}
\frac{\partial^2 \beta}{\partial t \partial y} = 0, 
\end{equation}
The result $\partial \beta / \partial y = 0$ means that $A=D$ from Eq.\
\eqref{c1set10} and $C=B$ from Eq. \eqref{c1set04}. Hence, the set of
equations from Eq.\,\eqref{c1set01} to Eq.\,\eqref{c1set09} simplify
for $\partial \beta /\partial y = 0$,
\begin{eqnarray}
&& \mu = \beta^2 (D - p) + \frac{D}{3}\,=\,\beta^2 (A - p) + \frac{A}{3},
\label{newc1set01}
\\ [2pt]
&& \frac{\partial}{\partial x}\left[ \frac{\partial \beta}{\partial t}
+ \frac{1}{2}\frac{\partial}{\partial x}(\beta^2) \right] = -64 \pi
B\,= -64 \pi C. 
\label{newc1set05}
\end{eqnarray}
From the result $\partial^2 \beta / \partial x \partial y = 0$ of Eq.\
\eqref{AAAA-A} the set of equations \eqref{c1set01} to \eqref{c1set10}
is recovered. Therefore, Eqs.\ \eqref{AAAA-A} show that this case
separates itself in two conditions, either $\partial\beta/\partial x=0$
or $\partial\beta/ \partial y = 0$. Next we analyze both conditions.

\item[Case 1a: \small $\bm{\left[\displaystyle \frac{\partial
     \beta}{\partial z} = 0 \,\,\,\, \text{and} \,\,\,\, 
		 \frac{\partial \beta}{\partial x} = 0\right]}$]
Setting Eq.\eqref{c1set09} equal to zero then $B = - C$, and the set of
equations \eqref{c1set01} to \eqref{c1set10} simplify to,
\begin{eqnarray} 
\mu &=& \beta^2 (2D - A - p) + \frac{A}{3},
\label{c1aset01}
\\ [2pt]
B &=& - C = \frac{1}{3}A,
\label{c1aset02}
\\ [2pt]
\left(\frac{\partial \beta}{\partial y}\right)^2 &=& 32 \pi C,
\label{c1aset03}
\\ [2pt]
\frac{\partial^2 \beta}{\partial y^2} &=& 16 \pi \beta (A-D).
\label{c1aset04}
\end{eqnarray}
For this case, there is an equation of state given by Eq.\
\eqref{c1aset01}, $\beta$ is a function of time and $y$ coordinates
and must be found by solving both Eqs.\,\eqref{c1aset03} and 
\eqref{c1aset04} in terms of the pressures $A, C$ and $D$. Note that
Eq.\ \eqref{c1aset02} relates the pressures $A, B$ and
$C$. The EMT for this case may be written as follows,
\begin{equation}
T_{\alpha \sigma} = 
\begin{pmatrix} 
\beta^2 (2D-A) + A/3 & - \beta D & 0  & 0  \\ 
- \beta D     & A         & 0  & 0  \\ 
0             & 0         & A/3  & 0  \\ 
0             & 0         & 0  & -A/3 
\end{pmatrix}\,.
\label{emtc1a}
\end{equation}
One should also note that for the $T_{00}$ component from Eq.\
\eqref{emtc1a} to be of positive value the following inequality
must hold,
\begin{equation}
\beta^2 > \frac{A}{3(A-2D)}.
\label{ineq1b}
\end{equation}

\item[Case 1b: \small $\bm{\left[\displaystyle \frac{\partial
     \beta}{\partial z} = 0 \,\,\,\, \text{and} \,\,\,\, 
     \frac{\partial \beta}{\partial y} = 0\right]}$]

Since Eq.\ \eqref{c1set04} is equal to zero, then $B = C$. Similarly
for Eq.\ \eqref{c1set10} it is clear that $A = D$. Consequently, the
set of equations \eqref{c1set01} to \eqref{c1set10} simplify to,
\begin{align} 
&\mu = \beta^2 (2D - A - p) + \frac{A}{3},
\label{c1bset01}
\\[2pt]
&B = C + \frac{2}{3}A,
\label{c1bset02}
\\[2pt]
&B = C,
\label{c1bset03}
\\[2pt]
&A = D,
\\[2pt]
&\frac{\partial}{\partial x}\left[ \frac{\partial \beta}{\partial t}
+ \frac{1}{2}\frac{\partial}{\partial x}(\beta^2) \right] = -64\pi B. 
\label{c1bset04}
\end{align}

But, from Eq. \eqref{c1set04} we have that $A=0$ because $B=C$. So,
from Eq. \eqref{newc1set01} we have that $\mu = -\beta^2 p$, and one
is left with a non homogeneous Burgers equation. This case reduces
the above set of equations to 
\begin{align} 
&\mu = - \beta^2 p,
\label{c1bfinal01}
\\[2pt]
&B = C,
\label{c1bfinal02}
\\[2pt]
&\frac{\partial}{\partial x}\left[
\frac{\partial \beta}{\partial t}
+ \frac{1}{2}\frac{\partial}{\partial x}(\beta^2)
\right]
= -64\pi B. 
\label{c1bfinal03}
\end{align}

Eq.\ \eqref{c1bfinal01} represents an equation of state between
matter density $\mu$ and the pressure $p$. The pressures $B$ and
$C$ are functions of the spacetime coordinates $(t,x,y,z)$ and
from Eq.\ \eqref{c1bfinal02} they are equal. Eq.\ \eqref{c1bfinal03}
is a non homogeneous Burgers equation, since its right hand side
cannot be readily integrated. It might mean that there is no
conservation law that can describe the warp drive for the PPF EMT.
The only possible way leading to conservation law is for $B$ being
a constant, which then allows a straightforward integration. However,
since this is not the case, namely, all the pressures from the EMT are
not, necessarily, constant functions, it is necessary to determine
this functions through the boundary conditions. The EMT for the Case
1b case then yields,
\begin{equation}
T_{\alpha \sigma} = 
\begin{pmatrix} 
0 & 0 & 0  & 0  \\ 
0     & 0  & 0  & 0  \\ 
0     & 0  & B  & 0  \\ 
0     & 0  & 0  & B 
\end{pmatrix}\,.
\label{emtc1b}
\end{equation}

The only non vanishing components of the PPF EMT above are $T_{22}$
and $T_{33}$. This case recovers the perfect fluid Case 1b, that is,
the dust EMTs when one chooses $p = B = 0$. For $B \neq 0$, one would
have to solve Eq.\ \eqref{c1bfinal03} to determine how the bubble
moves in this type of fluid spacetime. Again, negative matter density
emerges from the Einstein equation solutions for this specific choice
of EMT.

Nevertheless, this is a rather peculiar EMT, since the $T_{00}$
component is zero, but the equation of state \eqref{c1bfinal01}
remains and only two diagonal terms are not zero. In addition, it
is contradictory with the perfect fluid solution because the PPF
reduces to the perfect fluid under the condition
\begin{equation}
p=A=B=C=D,
\label{ppfredux}
\end{equation}
but in this solution $A=0$, but $B\not=0$. So, we discard this case
as unphysical.

\item[Case 2: \small $\bm{\left[\displaystyle\frac{\partial \beta}
     {\partial y}=0\right]}$]

The set of equations \eqref{set01} to \eqref{set10}
simplify to,
\begin{eqnarray}
&& \mu = \beta^2 (2D - A - p) + \frac{A}{3},
\label{c2set01}
\\ [2pt]
&& \frac{\partial^2 \beta}{\partial x \partial z} = 0,
\label{c2set02}
\\ [2pt]
&& \left(\frac{\partial \beta}{\partial z}\right)^2 = - \frac{32}{3}
\pi A = -\,16 \pi (B-C),
\label{c2set04}
\\ [2pt]
&& 2 \frac{\partial \beta}{\partial z} \, \frac{\partial \beta}
{\partial x} + \frac{\partial^2 \beta}{\partial t \partial z} = 0,
\label{c2set06}
\\ [2pt]
&& \frac{\partial}{\partial x}\left[ \frac{\partial \beta}{\partial t}
+ \frac{1}{2}\frac{\partial}{\partial x}(\beta^2) \right] = -32 \pi
(B + C)\,\,, 
\label{c2set09}
\\ [2pt]
&& \frac{\partial^2 \beta}{\partial z^2} = 16 \pi \beta (A-D).
\label{c2set10}
\end{eqnarray}

Eq.\ \eqref{c2set04} straightforwardly implies the following
relationship between pressures $A, B$ and $C$,
\begin{equation}
B = C + \frac{2}{3}A.
\end{equation}
Eqs.\ \eqref{c2set02} and \eqref{c2set04} indicate that $A$ and $B-C$
do not depend on the $x$-coordinate. In addition, for real solutions $A$
must be negative or zero. Finally, Eq.\ \eqref{c2set02} shows that that
this case also unfolds in two subcases: $\partial \beta/\partial z=0$ 
and/or $\partial \beta/\partial x=0$.  

\item[Case 2a: \small $\bm{\left[\displaystyle \frac{\partial
     \beta}{\partial y} = 0 \,\,\,\, \text{and} \,\,\,\, \frac{\partial
     \beta}{\partial x} = 0\right]}$]
		
Clearly $B = -C$ as a consequence of Eq.\ \eqref{c2set09}. Then, the
set of equations Eqs.\ \eqref{c2set01} to \eqref{c2set10} simplifies to,
\begin{eqnarray} 
&& \mu = \beta^2 (2D - A - p) + \frac{A}{3},
\label{c2aset01}
\\ [2pt]
&& B=-C=\frac{A}{3},
\label{c2aset02}
\\ [2pt]
&& \left(\frac{\partial \beta}{\partial z}\right)^2 = 32 \pi C,
\label{c2aset03}
\\ [2pt]
&& \frac{\partial^2 \beta}{\partial z^2} = 16 \pi \beta (A-D).
\label{c2aset04}
\end{eqnarray}

The EMT for this case is given by the following expression,
\begin{equation}
T_{\alpha \sigma} = 
\begin{pmatrix} 
\beta^2 (2D-A) + A/3 & - \beta D & 0  & 0  \\ 
- \beta D     & A         & 0  & 0  \\ 
0             & 0         & - A/3  & 0  \\ 
0             & 0         & 0  & A/3 
\end{pmatrix},
\label{emtc2a}
\end{equation}
which is almost equal to Case 1a (Eq.\ \ref{emtc1a}), apart from the
change of signs in the components $T_{22}$ and $T_{33}$. Both have
the same $T_{00}$ component and hence, the same equation of state,
and the condition for $T_{00}$ to be of positive is given by, 
\begin{equation}
\beta^2 > \frac{A}{3(A-2D)}.
\label{ineq1}
\end{equation}
Besides, this expressions determines another inequality that
$T_{00}$ must follow to be positive,
\begin{equation}
A - 2D > 0.
\label{ineq1c}
\end{equation}

\item[Case 2b: \small $\bm{\left[\displaystyle \frac{\partial
     \beta}{\partial y} = 0 \,\,\,\, \text{and} \,\,\,\, \frac{\partial
     \beta}{\partial z} = 0\right]}$]

The set of equations Eqs.\ \eqref{c2set01} to \eqref{c2set10}
immediately simplify to,
\begin{align} 
&\mu = \beta^2 (2D - A - p) + \frac{A}{3},
\label{c2bset01}
\\[2pt]
&B = C - \frac{2}{3}A,
\label{c2bset02}
\\[2pt]
&B = C,
\label{c2bset03}
\\[2pt]
&A = D = 0,
\\[2pt]
&\frac{\partial}{\partial x}\left[ \frac{\partial \beta}{\partial t}
+ \frac{1}{2}\frac{\partial}{\partial x}(\beta^2) \right] = -64\pi
B\,= -64\pi C\,\,. 
\label{c2bset04}
\end{align}
Therefore $\mu = -\beta^2 p$, a non homogeneous Burgers equation
is also present, and the expressions above are reduced to 
\begin{align} 
&\mu = - \beta^2 p,
\label{c2bfinal01}
\\[2pt]
&B = C,
\label{c2bfinal02}
\\[2pt]
&\frac{\partial}{\partial x}\left[ \frac{\partial \beta}{\partial t}
+ \frac{1}{2}\frac{\partial}{\partial x}(\beta^2) \right] = -64\pi B, 
\label{c2bfinal03}
\end{align}
which are the same results as obtained in Case 1b, as well as the EMT,
that is,
\pagebreak
\begin{equation}
T_{\alpha \sigma} = 
\begin{pmatrix} 
0 & 0 & 0  & 0  \\ 
0     & 0  & 0  & 0  \\ 
0     & 0  & B  & 0  \\ 
0     & 0  & 0  & B 
\end{pmatrix}\,.
\label{emtc2b}
\end{equation}
\end{description}
\vspace{-0.6cm} $\square$ \vspace{0.3cm}

Again, because this solution cannot fulfill the requirement given by
Eq.\ (\ref{ppfredux}), in fact it contradicts it since one cannot have
both $A=0$ and $B\not=0$, this solution is, similarly to Case 1b,
dismissed as unphysical.

Table \ref{table_summary} summarizes the Einstein equations solutions
obtained as a result of the PPF EMT in the Alcubierre warp drive
metric.
\begin{table}[ht!]
\begin{tabular}{| m{3cm} | m{3cm} | m{6.5cm} |}
\hline 
Case & Conditions & Results \\ 
\hline 
\multirow{2}{*}{$1) \
\displaystyle{\frac{\partial \beta}{\partial z} = 0}$}
& 
$1a) \ \displaystyle{\frac{\partial \beta}{\partial x} = 0}$
&
$\begin{array} {ll} 
\\ [-10pt]
\displaystyle{\mu = \beta^2(2D-A-p) + \frac{A}{3}} \\ [7pt]
\beta = \beta(t,y)\\ [7pt]
\displaystyle{B = - C = \frac{A}{3}}\\ [7pt]
\displaystyle{\left(\frac{\partial \beta}{\partial y}
\right)^2 = 32 \pi C},  \\ [10pt]
\displaystyle{\frac{\partial^2\beta}{\partial y^2} =
16 \pi \beta (A-D)}
\\[7pt]
\end{array}$ \\
\cline{2-3}   
&
$1b) \ \displaystyle{\frac{\partial \beta}{\partial y} = 0}$
&
$\begin{array} {ll} 
\\ [-10pt]
\displaystyle{\mu = - \beta^2p}\\ [7pt]
\beta = \beta(t,x)\\ [7pt]
\displaystyle{B = C}  \\ [5pt]
\displaystyle{A = D = 0} \\ [7pt]
\displaystyle{\frac{\partial}{\partial x}\left[
\frac{\partial \beta}{\partial t} 
+ \frac{1}{2} \frac{\partial}{\partial x}(\beta^2)
\right] = - 64 \pi B} \\[8pt]
\rightarrow\mbox{solution \textit{dismissed} as unphysical} \\
\end{array}$ \\
\hline

\multirow{2}{*}{$2) \
\displaystyle{\frac{\partial \beta}{\partial y} = 0}$}
& 
$2a) \ \displaystyle{\frac{\partial \beta}{\partial x} = 0}$
&
$\begin{array} {ll} 
\\ [-10pt]
\displaystyle{\mu = \beta^2(2D-A-p) + \frac{A}{3}} \\ [7pt]
\beta = \beta(t,z)\\ [7pt]
\displaystyle{B = - C = \frac{A}{3}}\\ [7pt]
\displaystyle{\left(\frac{\partial \beta}{\partial z}
\right)^2 = 32 \pi C}  \\ [9pt]
\displaystyle{\frac{\partial^2\beta}{\partial z^2} =
16 \pi \beta (A-D)}
\\[7pt]
\end{array}$ \\
\cline{2-3}   
&
$2b) \ \displaystyle{\frac{\partial \beta}{\partial z} = 0}$
&
$\begin{array} {ll} 
\\ [-10pt]
\displaystyle{\mu = - \beta^2p}\\ [7pt]
\beta = \beta(t,x)\\ [7pt]
\displaystyle{B = C}  \\ [5pt]
\displaystyle{A = D = 0} \\ [7pt]
\displaystyle{\frac{\partial}{\partial x}\left[
\frac{\partial \beta}{\partial t} 
+ \frac{1}{2} \frac{\partial}{\partial x}(\beta^2)
\right] = - 64 \pi B}
\\[8pt]
\rightarrow\mbox{solution \textit{dismissed} as unphysical} 
\end{array}$ \\  
\hline 

\end{tabular}
\caption{Summary of all solutions of the Einstein equations for the
	Alcubierre warp drive metric having the parametrized perfect
	fluid (PPF) EMT. The cases, conditions and their respective
	designations are the same as in Table \ref{tab1}}
\label{table_summary}
\end{table}

\subsection{Discussion}

As in the perfect fluid situations, Cases 1b and 2b are the same for
the PPF, producing the same equations of state, coordinate dependency
for the $\beta$ function and a non-homogeneous Burgers equation. As
stated above, only for a constant pressure $B$ that a conservation law
could possibly emerge from the Burgers equation (\ref{c2bfinal03}) that
is common to both cases. Otherwise this expression cannot be readily
integrated unless further boundary conditions are met (see Section
\ref{1b2b} below). Nevertheless, because the solutions of these two
cases cannot satisfy the requirement established by Eq.\ (\ref{ppfredux})
for reducing the PPF to the perfect fluid, they are dismissed as
unphysical. 

Regarding Cases 1a and 2a, the only physically plausible solutions
remaining for the PPF EMT, they must obey the same inequalities
(\ref{ineq1}) and (\ref{ineq1c}) so that the $T_{00}$ component be
positive in both Eqs.\ (\ref{emtc1a}) and (\ref{emtc2a}). The
function $\beta$ depends on the $y$ coordinate in the former case
and the $z$ one in the later, and, respectively, are a result of the
integration of the following differential equations,
\begin{equation}
\beta\frac{\partial\beta}{\partial y}=\frac{C}{A-D},
\label{abc1}
\end{equation}
\begin{equation}
\beta\frac{\partial\beta}{\partial z}=\frac{C}{A-D}.
\label{abc2}
\end{equation}
However, the quantities $A$, $C$ and $D$ are function of the
coordinates, thus integrating the expressions above require further
conditions, such as the ones respectively discussed in Sections
\ref{sit1a} and \ref{sit2a} below.

\subsubsection{Equations of state}

The perfect fluid solutions have equations of state given by
\cite[Ref.][p.\ 14]{EllisElst},
\be
p=p(\mu)=(\gamma-1)\mu, \;\;\; \dot{\gamma}=\frac{\mathrm{d}\gamma}
{\mathrm{d}t}=0, 
\label{eq-state}
\ee
Ordinary fluids can be approximated with $1\leq\gamma\leq 2$, where
incoherent matter, or dust, corresponds to $\gamma = 1$, and radiation
to $\gamma=\frac{4}{3}$.

In the analyzes above we found that for the perfect fluid content
solution Cases 1a and 2a for the warp drive metric the equation of
state is given by $p = 3 \mu$ (see Table \ref{tab1}), which
corresponds to $\gamma = 4$. However, they do not allow a simple
expressions for the equation of state similar to Eq.\ (\ref{eq-state})
for the PPF solutions 1a and 2a (see Table \ref{table_summary}). The
PPF content solution for the Cases 1b and 2b were dismissed as
unphysical,


\section{EMT Divergence of the perfect fluid and PPF}\label{divemts}
\renewcommand{\theequation}{5.\arabic{equation}}
\setcounter{equation}{0}

In this section we will investigate the associated conservation laws
to the Einstein equations under a warp drive spacetime by means of the
usual condition that the EMT divergence must be zero for both the
perfect fluid and PPF. We shall start with the EMT for the PPF because
from its very definition the perfect fluid can be recovered by setting
the equality (\ref{ppfredux}).

For the fluids discussed here, setting ${T^{\alpha \sigma}}_{;\sigma}=0$
in the EMT (\ref{71}) results in the following expressions, 
\begin{align} 
\nonumber
- \frac{\partial \beta}{\partial x} (D + \mu) 
- \frac{\partial \mu}{\partial t} 
- \beta \left[\frac{\partial D}{\partial x} 
+ \frac{\partial \mu}{\partial x} 
+ \frac{\partial \beta}{\partial t}(2p + A - 3D)\right]& \\ 
+ \beta^2\left[\frac{\partial D}{\partial t} 
- \frac{\partial p}{\partial t} 
+ 3\frac{\partial \beta}{\partial x}(D-p)\right]
+ \beta^3 \left(\frac{\partial D}{\partial x} 
- \frac{\partial p}{\partial x}\right)&= 0,
\label{divT0}
\\[2pt]
\frac{\partial A}{\partial x} 
+ \frac{\partial \beta}{\partial t} (D-A)
+ \beta \left[ 3 \frac{\partial \beta}{\partial x} (D-A)
+ \frac{\partial D}{\partial t} 
- \frac{\partial A}{\partial t}\right] 
+ \beta^2 \left(\frac{\partial D}{\partial x}
- \frac{\partial A}{\partial x}\right) &= 0,
\label{divT1} 
\\[2pt]
\frac{\partial B}{\partial y} +
\beta \frac{\partial \beta}{\partial y} (D-A) &= 0,
\label{divT2}
\\[2pt]
\frac{\partial C}{\partial z} +
\beta \frac{\partial \beta}{\partial z} (D-A) &= 0.
\label{divT3}
\end{align}

\subsection{Case 1a: $\bm{\left[\displaystyle\frac{\partial \beta}
{\partial z} = 0\right.}$ and $\bm{\left.\displaystyle\frac{\partial
\beta}{\partial x} = 0\right]}$}\label{sit1a}

Eqs.\ \eqref{divT0} to Eq.\,\eqref{divT3} are reduced to the ones below, 
\begin{align}
\nonumber
- \frac{\partial \mu}{\partial t} - \beta \left[\frac{\partial D}
{\partial x}+\frac{\partial \mu}{\partial x}+\frac{\partial \beta}
{\partial t}(2p + A - 3D)\right]+\beta^2 \left(\frac{\partial D}
{\partial t}-\frac{\partial p}{\partial t}\right)& \\
+ \beta^3 \left(\frac{\partial D}{\partial x} - \frac{\partial p}
{\partial x}\right)&= 0,
\label{npf1adivT0}
\\[2pt]
\frac{\partial A}{\partial x}+\frac{\partial \beta}{\partial t}(D-A)
+ \beta \left(\frac{\partial D}{\partial t}-\frac{\partial A}
{\partial t} \right)+\beta^2 \left(\frac{\partial D}{\partial x}
- \frac{\partial A}{\partial x}\right)&= 0, \label{npf1adivT1} 
\end{align}
%
\begin{align}
\frac{\partial B}{\partial y} + \beta \frac{\partial \beta}
{\partial y} (D-A)&= 0,
\label{npf1adivT2}
\\[2pt]
\frac{\partial C}{\partial z}&= 0.
\label{npf1adivT3}
\end{align}
The results above concern the PPF. These four equations together with
the five ones shown in the respective results of Table
\ref{table_summary} means an overdetermined system for the six 
unknowns $\mu, p, A, B, C$, $D$.

The perfect fluid is recovered by setting Eq.\ (\ref{ppfredux}).
Hence, Eqs.\ \eqref{npf1adivT0} to \eqref{npf1adivT3} become,
\begin{align}
- \frac{\partial \mu}{\partial t}-\beta \left(\frac{\partial p}{\partial x}
+ \frac{\partial\mu}{\partial x}\right)&= 0,
\label{pf1adivT0}
\\[2pt]
\frac{\partial p}{\partial x}&= 0,
\label{pf1adivT1} 
\\[2pt]
\frac{\partial p}{\partial y}&= 0,
\label{pf1adivT2}
\\[2pt]
\frac{\partial p}{\partial z}&= 0,
\label{pf1adivT3}
\end{align}
and, thus, the pressure does not depend on the spatial coordinates.
In addition, Eq.\ \eqref{pf1adivT0} reduces to the expression below, 
\begin{equation}
\frac{\partial \mu}{\partial t} 
+ \beta \frac{\partial \mu}{\partial x} = 0,
\label{divpf0}
\end{equation} 
which is the continuity equation, where $\mu$ plays the role of the
fluid density and $\beta$ is the flow velocity vector field. It is worth
mentioning that for a constant density the fluid has incompressible
flow, then all the partial derivatives of $\beta$ in terms of the
spatial coordinates vanish and the flow velocity vector field has null
divergence, this being a classical fluid dynamics scenario, and the local
volume  dilation rate is zero.

\subsection{Case 2a: $\bm{\left[\displaystyle\frac{\partial \beta}
{\partial y} = 0\right.}$ and $\bm{\left.\displaystyle\frac{\partial
\beta}{\partial x} = 0\right]}$}\label{sit2a}

Equations \eqref{divT0} to \eqref{divT3} simplify to the following
expressions:
\begin{align}
\nonumber
- \frac{\partial \mu}{\partial t}-\beta \left[\frac{\partial D}
{\partial x}+\frac{\partial \mu}{\partial x}+\frac{\partial \beta}
{\partial t}(2p + A - 3D)\right]&
\\
+ \beta^2\left(\frac{\partial D}{\partial t}-\frac{\partial p}
{\partial t}\right)+\beta^3 \left(\frac{\partial D}{\partial x} 
- \frac{\partial p}{\partial x}\right)&= 0,
\label{npf2adivT0}
\\[2pt]
\frac{\partial A}{\partial x}+\frac{\partial \beta}{\partial t}(D-A)
+ \beta \left(\frac{\partial D}{\partial t}-\frac{\partial A}
{\partial t} \right)+\beta^2 \left(\frac{\partial D}{\partial x}
- \frac{\partial A}{\partial x}\right)&= 0,
\label{npf2adivT1} 
\\[2pt]
\frac{\partial B}{\partial y}&= 0,
\label{npf2adivT2}
\\[2pt]
\frac{\partial C}{\partial z}+\beta \frac{\partial \beta}
{\partial z} (D-A)&= 0.
\label{npf2adivT3}
\end{align}
Assuming Eq.\ (\ref{ppfredux}) the perfect fluid is recovered, resulting
in the already discussed Eqs.\ \eqref{pf1adivT0} to \eqref{pf1adivT3},
as well as the continuity equation \eqref{divpf0}. 

\subsection{Cases 1b and 2b: $\bm{\left[\displaystyle\frac{\partial \beta}
{\partial y} = 0\right.}$ and $\bm{\left.\displaystyle\frac{\partial
\beta}{\partial z} = 0\right]}$}\label{1b2b}

It follows from Table \ref{tab1} that conditions 1b and 2b for the
perfect fluid imply $\mu = p = 0$, which also means $A=B=C=D=0$. Then,
Eqs.\ (\ref{divT0}) to (\ref{divT3}) of the null EMT divergence are
are immediately satisfied for the perfect fluid.

Regarding the PPF, we have already seen that in these cases the
solutions were dismissed as unphysical. Nonetheless, it is worth
analyzing the resulting expressions to show that they lead to trivial
cases or to the dust solution already studied in Ref.\ \cite{nos}.

Table \ref{table_summary} shows the following solutions for the PPF
EMT considering Cases 1b and 2b: $A=D=0$, $B=C$, and $\mu=-\beta^2 p$.
Hence, Eqs.\ (\ref{divT0}) to (\ref{divT3}) are reduced to the
following expressions,
\begin{align}
-\frac{\partial \mu}{\partial t}-\beta \left(\frac{\partial \mu}
{\partial x}+2p \frac{\partial \beta}{\partial t}\right)-\beta^2
\frac{\partial p}{\partial t}-\beta^3 \frac{\partial p}{\partial x}
&= 0,
\label{newdivT0}
\\[2pt]
\frac{\partial B}{\partial y}&= 0,
\label{newdivT2}
\\[2pt]
\frac{\partial C}{\partial z}=\frac{\partial B}{\partial z}&= 0.
\label{newdivT3}
\end{align}
Eqs.\ \eqref{newdivT2} and \eqref{newdivT3} imply that $B$ does not
depend on both the $y$ and $z$ coordinates. Inserting the state
equation $\mu=-\beta^2 p$ (see Table \ref{table_summary}) into Eq.\
\eqref{newdivT0} yields
\begin{equation}
p \beta^2 \frac{\partial \beta}{\partial x} = 0.
\end{equation}
This expression can be separated in three subcases: $\beta = 0$,
$p = 0$, $\partial \beta/\partial x = 0$. According to Eq.\
\eqref{c2bfinal03} the last subcase implies $B = 0$ and, hence,
there is no Burgers equation. Let us discuss below the consequences
of each of these subcases.

\subsubsection{{Subcase} $\left[\,\beta=0\,\right]$}

This solution reduces the warp drive metric \eqref{alcmetric1} into
a Minkowski flat spacetime. There is no warp bubble and, so, no warp
drive.
 
\subsubsection{{Subcase} $\left[\,B=0\,\right]$}

Remembering Eqs.\ (\ref{emtc1b}) and (\ref{emtc2b}) this means that
all EMT components are zero, even though, the equation of state
$\mu = - \beta^2 p$ remains. But, since there is no EMT both $\mu$
and $p$ must vanish, hence this case reduces itself to the dust
solution \cite{nos}.
 
\subsubsection{{Subcase} $\left[\,p=0\,\right]$}

Considering the equation of state $\mu = - \beta^2 p$ then the
matter density is also zero. The EMT for the PPF assumes the
form given by Eq.\ (\ref{emtc2b}), which implies in pressure
without matter density, an unphysical situation that should
either be dismissed or assumed to be just the dust solution.


\section{Energy conditions} \label{engconds}
\renewcommand{\theequation}{6.\arabic{equation}}
\setcounter{equation}{0}

The energy conditions are well known inequalities discussed by Hawking
and Ellis \cite{HawkingEllis1973} that may be applied to the 
matter content in physical systems as boundary conditions and to test
if the energy of such systems follows positive constraint values.

This section aims at obtaining these conditions for the Alcubierre
warp drive geometry considering the previously discussed EMTs for both
the perfect fluid and PPF. Our focus will be on the main classical
inequalities, namely, the \textit{weak}, \textit{dominant},
\textit{strong} and \textit{null energy conditions}. Our analysis
starts with the PPF EMT defined in Eq.\ \eqref{71} in order to
constrain its quantities so that the inequalities are satisfied for
each of the four just named conditions. Then the same analysis for
the perfect fluid is performed by reducing the results to the choice
set by Eq.\ (\ref{ppfredux}), but considering only the physically
plausible cases.

\subsection{Weak Energy Condition (WEC)}\label{wec-cond} 

The WEC requires that at each point of the spacetime the EMT must
obey the following inequality, 
\begin{equation}
T_{\alpha \sigma} \, u^\alpha u^\sigma \geq 0.    
\label{wec111}
\end{equation}
This is valid for any timelike vector $\textbf{u} \, (u_\alpha u^\alpha
< 0)$ as well as any null vector $\textbf{k} \, (k_\alpha k^\alpha=0$;
see Section \ref{neccondi} below). For an observer with a unit tangent
vector $\textbf{v}$ at some point of the spacetime, the local energy
density measured by any observer is non-negative \cite{HawkingEllis1973}.
Considering the Eulerian (normal) observers from \cite{Alcubierre1994}
with 4-velocity given by the expressions below
\begin{equation}
u^\alpha = (1, \beta, 0, 0)\, , \ \ 
u_\alpha = (- 1,0,0,0),
\end{equation}
and the EMT, given by Eq.\,\eqref{71}, computing the expression
$T_{\alpha \sigma}\,u^\alpha u^\sigma$ for the WEC yields,
\begin{align}
\nonumber T_{\alpha \sigma} \, u^\alpha u^\sigma &= 
T_{00} \, u^0 u^0 + 2 T_{01} \, u^0 u^1 + T_{11} \, u^1 u^1 \\
&= \mu + \beta^2(p - 2 D + A) \,. 
\label{eq88}
\end{align}
Let us now calculate the WEC for all perfect fluid and PPF Einstein
equations solutions obtained in the previous sections as referred to
in Table \ref{table_summary}.

\subsubsection{Cases 1a and 2a}

Let us start by considering the PPF. Substituting the equation of
state \eqref{matterdens} into Eq.\ \eqref{eq88} yields, 
\begin{equation}
T_{\alpha\sigma}\, u^\alpha u^\sigma=\frac{A}{3} \geq 0.
\label{wec}
\end{equation}
So, the pressure $A$ must positive. Considering that both cases led
to Eq.\ (\ref{c2aset02}), the pressure $C$ must be negative. Besides,
for the state equation (\ref{matterdens}) the
density $\mu$ becomes positive if 
\be
A+p\leq 2D\,\,.
\ee
This inequality implies that it is possible for the fluid density
to account for both negative and positive values in a local way
depending on how the momentum and pressure components relate to
each other in the spacetime.

Remembering Eq.\ (\ref{ppfredux}), the resulting inequality (\ref{wec})
indicates that the WEC is satisfied with the perfect fluid EMT for
a positive pressure $p$. Considering Eqs.\ (\ref{p3mu}) and (\ref{bbb1})
that would mean a complex solution for $\beta$, possibility already
envisaged in Section \ref{discu}. Note that this is a general result
for the perfect fluid with a cross term solution.

\subsubsection{Cases 1b and 2b}

It has been shown above that in these two cases the PPF solutions
result in $A=D=0$ and $\mu=-\beta^2 p$, which substituted in Eq.\
(\ref{eq88}) yield  $T_{\alpha \sigma}\,u^\alpha u^\sigma=0$. Hence,
the WEC is not violated for the PPF.

For the perfect fluid, Table \ref{tab1} shows that in these two cases
the perfect fluid solutions reduce to the dust content solution for
the warp drive metric, which trivially satisfies the WEC \cite[Ref.]
[p.\ 6]{nos}.

\subsection{Dominant Energy Condition (DEC)}\label{domecon}

The DEC states that for every timelike vector $u_a$ the following
inequality must be satisfied,
\begin{equation}
T^{\alpha \beta} \, u_\alpha u_\beta \geq 0, \quad \text{and} \quad 
F^\alpha  F_\alpha  \leq 0, 
\end{equation}
where $F^\alpha = T^{\alpha \beta} u_\beta$ is a non-spacelike vector.
This condition means that for any observer the local energy density is
non-negative and the local energy flow vector is non-spacelike. In any
orthonormal basis the energy dominates the other components of the EMT, 
\begin{equation}
T^{00} \geq |T^{ab}|, \ \text{for each} \ a, b.
\end{equation}
Hawking and Ellis \cite{HawkingEllis1973} suggested that this condition
must hold for all known forms of matter and that it should be the case
in all situations.

Evaluating this condition for PPF, the first DEC inequality is given
by the following expression,
\be
T^{\alpha \beta} \, u_\alpha u_\beta = 
T^{00} = \beta^2(A - 2D + p) + \mu \geq 0,
\ee 
whereas the second inequality $F^\alpha F_\alpha \leq 0$ yields,
\begin{align}
\nonumber 
(T^{\alpha \beta} \, u_\beta) (T_{\alpha \beta} \, u^\beta)
=& - \mu^2 - A^2 \beta^4 - \beta^4 p^2 + A^2 \beta^2 
- (4 \beta^4 - \beta^2) D^2\nonumber \\
&+2 (2 A \beta^4 - A \beta^2) D - 2 (A \beta^2 - 2 \beta^2 D)
\mu \nonumber\\ 
&-2 (A \beta^4 - 2 \beta^4 D + \beta^2 \mu) p \leq 0.
\end{align}
As before, let us analyze next the solutions of the Einstein equations
considering each group of subcases as referred to in Table
\ref{table_summary}.

\subsubsection{Cases 1a and 2a}

The two important results for these two cases are
Eqs.\ (\ref{c2aset01}) and (\ref{c2aset02}). Substituting them into
the two DEC inequalities above result in the following expressions,
\begin{align}
T^{\alpha \beta} \, u_\alpha u_\beta&= \frac{A}{3} \geq 0,
\label{dec1}
\\
(T^{\alpha \beta} \, u_\beta) (T_{\alpha \beta} \, u^\beta)&=
\left[(A-D)\beta+\frac{A}{3}\right]\left[(A-D)\beta-\frac{A}{3}
\right] \leq 0.
\label{dec2}
\end{align}
Eq.\ \eqref{dec1} is equal to Eq.\ (\ref{wec}), so this dominant
energy condition is the same as the weak one. Eq.\ \eqref{dec2}
separates in two inequalities,
\begin{align}
&-\frac{A}{3\left(D-A\right)}\leq\beta\leq\frac{A}{3\left(D-A\right)},
\label{dec3}
\\
&-\frac{A}{3\left(A-D\right)}\leq\beta\leq\frac{A}{3\left(A-D\right)},
\label{dec4}
\end{align}
which can be rewritten as
\be
\left|\beta\right| \leq \frac{1}{3} \left|\frac{A}{D-A}\right|.
\label{dec5}
\ee
In addition to the inequalities above, Eq.\ (\ref{dec2}) also provides
the following solutions,
\begin{equation}
\beta=\pm\frac{A}{3(A-D)}.
\label{dec5aa}
\end{equation}

Equation \eqref{dec5} shows that $\beta$ is upper bounded, but 
since both $A$ and $D$ have no restraints,
$\beta$ can be greater than unity. Remembering that $\beta =
v_s(t) f[r_s(t)]$, the dominant energy condition is not violated
for cases where the apparent warp bubble speed is greater than
the speed of light. This result also depends on the relation
between the pressure component $A$ and the momentum component
$D$ as can be seen in Eqs.\ \eqref{dec3} and \,\eqref{dec4}. As
an example suppose that $A=D+1$ in Eq.\ \eqref{dec3}, so that
\be
- \frac{D+1}{3} \leq v_s(t) f[r_s(t)] \leq \frac{D+1}{3}.
\ee
The symmetry on the negative and positive values that $\beta$ may
assume means the direction in which the warp bubble is moving on
the $x$-axis on the spatial hyper surface. Note that, depending
on the value of $D$, the warp bubble may assume the speed $v_s(t)$
greater than the speed of light.

Regarding the perfect fluid, choosing as in Eq.\ (\ref{ppfredux})
the DEC is given by the following expressions,
\begin{align}
\label{4.17-4.18}
T^{\alpha \beta} \, u_\alpha u_\beta =\frac{p}{3}&\geq 0, \\[4pt]
(T^{\alpha \beta} \, u_\beta)(T_{\alpha\beta} \, u^\beta) =p^2&\geq 0,
\end{align}
which means that for the DEC to be satisfied a positively valued matter
density is enough (see Table \ref{tab1}), although this also means
a complex result for $\beta$ (see Section \ref{discu}).

\subsubsection{Cases 1b and 2b}

In these cases the PPF solutions, although considered unphysical,
produced $B=C$, $A=D=0$ and $\mu=-\beta^2 p$ (see Table
\ref{table_summary}). So, as shown by the expressions below the DEC
is immediately satisfied, 
\begin{align}
&T^{\alpha \beta} \, u_\alpha u_\beta = T^{00} = 0, \\
&(T^{\alpha \beta} \, u_\beta) (T_{\alpha \beta} \, u^\beta)
= 0\,\,,
\end{align}
The perfect fluid is the dust solution for both cases (see Table
\ref{tab1}), which is a vacuum solution and satisfies the
inequalities above trivially. 

\subsection{Strong Energy Condition (SEC)}\label{seccondi}

For any timelike vector $u^\alpha$ the EMT must obey the following
inequality for the SEC be valid,
\begin{equation}
\left(T_{\alpha \beta} - \frac{1}{2}T \, g_{\alpha \beta} \right) 
u^\alpha u^\beta \geq 0,
\end{equation}
This requirement is stronger than the WEC and only makes sense in
general relativistic framework because this theory is governed by the
Einstein equations. These conditions imply that gravity is always
attractive.

To obtain expression for the SEC let us start with the scalar
$T = g^{_{\alpha \beta}}T_{\alpha \beta}$, where $g^{\alpha \beta}$
is given by the Alcubierre warp drive metric and $T_{\alpha\beta}$
is the EMT for the PPF (Eq.\ \ref{71}). The result is shown below,
\begin{equation}
T = g^{_{\alpha \beta}}T_{\alpha \beta} = - \mu 
+ A + B + C + \beta^2(2 D - p - A).    
\label{812}
\end{equation}
Substituting the equation of state \eqref{matterdens} into
the expression above yields, 
\begin{equation}
T = \frac{2}{3}A + B + C.
\label{813}
\end{equation}
As shown in Section \ref{wec-cond} above, Eq.\ (\ref{matterdens})
implies that $T_{\alpha\sigma}u^\alpha u^\sigma=A/3$ for $u^\sigma=
(1,\beta,0,0)$. Inasmuch as $g_{\alpha\sigma} u^\alpha u^\sigma=-1$,
then the SEC are given by the following expression,
\begin{equation}
\left(T_{\alpha\sigma}-\frac{1}{2}T \, g_{\alpha\sigma}\right) 
u^\alpha u^\sigma = \frac{2A}{3} + \frac{(B+C)}{2}  \geq 0.
\label{secaaa}
\end{equation}
The particular cases, as summarized in Table \ref{table_summary}, can
now be discussed.
 
\subsubsection{Cases 1a and 2a}

These cases resulted in the constraint $B=-\,C=A/3$. Eq.\
(\ref{secaaa}) may then be rewritten in the form below,
\begin{equation}
\left(T_{\alpha \sigma} - \frac{1}{2}T \, g_{\alpha \sigma} \right) 
u^\alpha u^\sigma = \frac{2A}{3}=2B  \geq 0\,\,.
\end{equation}
Clearly, for the SEC be satisfied it is only necessary that the
pressure component $A$ must be non negative.

Regarding the perfect fluid its respective SEC reads as,
\begin{equation}
\left(T_{\alpha \sigma} - \frac{1}{2}T \, g_{\alpha \sigma} \right) 
u^\alpha u^\sigma = \frac{2p}{3}  \geq 0.
\end{equation}
So, the SEC can be satisfied by the perfect fluid in these subcases
by requiring a positive pressure $p$.

\subsubsection{Cases 1b and 2b}

In these cases the PPF solutions stated that $A=D=0$ and $B=C$.
Hence, the SEC satisfied under the constraint below,
\begin{equation}
\left(T_{\alpha \sigma} 
- \frac{1}{2} T \, g_{\alpha \sigma}\right) 
u^\alpha u^\sigma = B \geq 0, 
\end{equation}
despite providing unphysical solutions. The perfect fluid solutions
are just the dust solution that is a vacuum solution \cite{nos}, so
satisfying the SEC trivially.

\subsection{Null Energy Condition (NEC)}\label{neccondi}

The SEC and WEC are satisfied in the limit of the null observers with
4-velocities $\textbf{k}$. To satisfy the NEC the EMT must follow
the inequality below, 
\begin{equation}
T_{\alpha \sigma} \, k^\alpha k^\sigma \geq 0, \ \text{for any 
null vector} \ k^\alpha.
\label{neceqcondi}
\end{equation}
To calculate the NEC let us suppose the following null vector
below,
\begin{equation}
k^\alpha = (a,b,0,0),    
\label{neck1}
\end{equation}
where the relation between $a$ and $b$ can be obtained by imposing 
the condition
\be
k_\alpha k^\alpha = 0,
\label{neck2}
\ee
that leads to
\be
a^2 \beta^2 - 2 a b \beta - a^2 + b^2 = 0,
\label{neck3}
\ee
whose roots for $a/b$ yield,
\begin{equation}
\frac{a}{b} = \frac{1}{\beta + 1} \ \ \text{and} \ \
\frac{a}{b} = \frac{1}{\beta - 1}.
\label{necroots}
\end{equation}
Considering the results above, the NEC may be written as follows, 
\begin{equation}
T_{\alpha \sigma} \, k^\alpha k^\sigma =
T_{00} k^0 k^0 + 2 T_{01} k^0 k^1 + T_{11} k^1 k^1,
\end{equation}
where $T_{\alpha \sigma}$ is the PPF EMT of Eq.\ \eqref{71}. So, the 
null condition is given by the following expression,
\begin{equation}
T_{\alpha \sigma} \, k^\alpha k^\sigma =
a^2 \beta^2 p - 2 a b \beta D + b^2 A + a^2 \mu \geq 0,
\label{820}
\end{equation}
Substituting the two roots in Eqs.\ \eqref{necroots} into Eq.\
\eqref{820} results in expressions below,
\begin{align}
&\frac{b^2}{(\beta + 1)^2} \, \left[\mu - \beta^2(2D - A - p) + 2\beta(A-D) 
+ A\right] \geq 0,
\label{necin1}
\\
&\frac{b^2}{(\beta - 1)^2} \, \left[\mu - \beta^2(2D - A - p) - 2\beta(A-D) 
+ A\right] \geq 0.
\label{necin2}
\end{align}

Next, as previously, we shall proceed to investigate first the Cases 1a
and 2a and then 1b and 2b ones, as defined in Tables \ref{tab1} and
\ref{table_summary}, for the PPF and the perfect fluid.

\subsubsection{Cases 1a and 2a}

Considering the PPF first, the equation of state (\ref{c2aset01}) is a
solution of both cases (see Table \ref{table_summary}), so the
resulting NEC given by Eqs.\ \eqref{necin1} and \eqref{necin2} simplify
to the following expressions,
\begin{align}
&\frac{b^2}{(\beta+1)^2}\,\left[\frac{4A}{3}+2\beta(A-D)\right]\geq 0,
\label{nec1}
\\
&\frac{b^2}{(\beta-1)^2}\,\left[\frac{4A}{3}-2\beta(A-D)\right]\geq 0,
\label{nec2}
\end{align}
whose solutions for $\beta$ yield the NEC for the PPF EMT, as below,
\begin{align}
&\beta\geq\frac{2}{3}\frac{A}{D-A} \ \ \text{and} \ \ \beta\neq - 1,
\label{null1}
\\
&\beta\leq \frac{2}{3}\frac{A}{A-D} \ \ \text{and} \ \ \beta\neq 1.
\label{null2}
\end{align}

Remembering that $\beta=v_s(t)f(r_s)$, then Eq.\ \eqref{null1} states
that the warp bubble can not assume the negative sign of the speed of
light, but it may assume values above it. To verify that, we have to
choose $A=D-1>2/3$ and $\beta>1$. Eq.\ \eqref{null2} states that $\beta$
cannot assume the exact value of the speed of light and it is bounded
by a superior value which can be greater than the speed of light. To
have this, one just have to choose for example $A = D + 1 > 2/3$.

For the perfect fluid, according to Eq.\ (\ref{ppfredux}) all pressures
are equal. Hence, substituting both values for $a/b$ found in Eq.\
\eqref{820} into Eqs.\ \eqref{necin1} and \eqref{necin2} they then
simplify to expressions below,
\begin{align}
&\frac{b^2}{(\beta + 1)^2} \, \left(\mu + p\right) \geq 0,
\\
&\frac{b^2}{(\beta - 1)^2} \, \left(\mu + p\right) \geq 0.
\end{align}
These results mean that for the NEC is satisfied for the perfect
fluid provided that
\be
\mu + p \geq 0.
\ee
Remembering that $p=3\mu$ for the perfect fluid in theses two cases,
then the NEC is fulfilled if the matter density does not have
negative values.

\subsubsection{Cases 1b and 2b}

These cases, although set as unphysical, we already concluded that
the PPF relates its pressures by the following expressions: $A=D=0$
and $B=C$. Therefore, Eqs.\ \eqref{necin1} and \eqref{necin2} yield,
\begin{align}
&\frac{b^2}{(\beta + 1)^2} \, \left(\mu + \beta^2 p \right) 
\geq 0,
\\
&\frac{b^2}{(\beta - 1)^2} \, \left(\mu + \beta^2 p \right) 
\geq 0.
\end{align}
So, the NEC establishes that $\mu+\beta^2 p\geq 0$, which is readily
satisfied since in these cases we already concluded that $\mu=-\beta^2
p$, provided that $\beta\neq-1$ for the former and $\beta\neq 1$ for
the latter. Regarding the perfect fluid, these cases reduce to the dust
solution studied in \cite{nos}, which is a vacuum solution and, therefore,
immediately satisfies the NEC.

Table \ref{tab3} summarizes the results of all energy conditions for
the PPF and perfect fluid EMTs with the Alcubierre warp drive metric.

\begin{table}[hb!]
\begin{tabular}{| m{5cm} | m{2.3cm} | m{8.2cm} |}
\hline 
Cases (refer to Tables \ref{tab1} and \ref{table_summary})
&
Energy Conditions
&
Results \\ 
\hline 
\multirow{4}{*}{1a and 2a} 
& 
Weak
&
$\begin{array} {ll} 
\\[-6pt]
\text{Perfect fluid:} \ 
\displaystyle{p \geq 0 }
\\[8pt]
\text{PPF:} \
\displaystyle{A \geq 0 }
\\[8pt]
\end{array}$ \\[8pt]
\cline{2-3}   
& 
Dominant
&
$\begin{array} {ll} 
\\[-6pt]
\text{Perfect Fluid:} \ \displaystyle{p \geq 0} \\[8pt]
\text{PPF:} \ \displaystyle{A \geq 0} \ \ \text{and} \ \ 
\displaystyle{\left|\beta\right| \leq \frac{1}{3}
\left|\frac{A}{D-A}\right|} \\[12pt]
\end{array}$ 
\\[8pt]
\cline{2-3}   
&
Strong
&
$\begin{array} {ll} 
\\[-6pt]
\text{Perfect fluid}: \ 
\displaystyle{p} \geq 0 
\\[8pt]
\text{PPF}: \
\displaystyle{A} \geq 0 
\\[8pt]
\end{array}$ 
\\[8pt]
\cline{2-3} 
&
Null
&
$\begin{array} {ll} 
\\[-6pt]
\text{Perfect fluid}: \
\displaystyle{ \mu + p \geq 0} 
\\[8pt]
\text{PPF}: \
\displaystyle{\beta \geq \frac{2}{3} \frac{A}{D - A} \ \ \text{and}
\ \ \beta \neq - 1} 
\\[8pt]
\text{PPF}: \
\displaystyle{\beta \leq \frac{2}{3} \frac{A}{A - D} \ \ \text{and}
\ \ \beta \neq 1} 
\\[8pt]
\end{array}$ 
\\[8pt]
\hline
\multirow{4}{*}{1b and 2b} 
& 
Weak
&
\text{immediately satisfied for perfect fluid and PPF.} 
\\ [2pt]
\cline{2-3}   
&
Dominant
&
\text{immediately satisfied for perfect fluid and PPF.}
\\ [2pt]
\cline{2-3}   
&
Strong
&
$\begin{array} {ll}
\\[-6pt]
\text{Perfect fluid: immediately satisfied.} \\ [8pt]
\text{PPF:} \ B\geq 0  \\ [2pt]
\end{array}$
\\[8pt]
\cline{2-3} 
&
Null
&
\text{immediately satisfied for perfect fluid and PPF.} \\ [2pt]
\cline{2-3}
\hline
\end{tabular}
\caption{Summary of all energy conditions results for the perfect
	fluid and PPF EMTs with the Alcubierre warp drive spacetime
        geometry.}
\label{tab3}
\end{table}


\section{Further discussions}\label{furdisc}
\renewcommand{\theequation}{7.\arabic{equation}}
\setcounter{equation}{0}

This section discusses some points, and raises others, all related to
the physics of the warp drive as suggested by the results presented
in the previous sections. It aims at offering some thoughts that may
be important in fostering further understanding on how a superluminal
travel can be achieved.

\subsection{Regulating Function and the Burgers Equation}

The regulating function {(\ref{regfunction})} describes the shape of the
warp bubble, but it is not uniquely determined. However, the integration
the Einstein equations in both the perfect fluid and the PPF led to the
appearance of generic functions in the dynamic equations which may end
up connected to the regulating function, a situation that adds to its
nonuniqueness. This means that physically feasible superluminal speeds
will require the specification of these generic functions, possibly by
boundary conditions. We shall show below an example of this situation
using the dust solution. 

Considering Eqs.\ (\protect\ref{betax}) and (\pr\ref{betax2}) the
partial derivative of the Burgers equation (\ref{pfadvec4}) yields,
\begin{equation}
\frac{\partial \beta}{\partial t} = f \frac{d^2 x_s}{dt^2}
+ v_s \frac{\partial f}{\partial r_s}\frac{\partial r_s}{\partial t},
\end{equation}
where the simplified notation $f=f\left[r_s(t)\right]$ and
$v_s=v_s(t)$ was adopted. Since,
\begin{equation}
\frac{\partial \beta}{\partial x} = v_s \frac{\partial f}{\partial r_s}
\frac{\partial r_s}{\partial x},
\end{equation}
the Burgers equation becomes,
\begin{equation}
f \frac{d^2 x_s}{dt^2}
+ v_s \frac{\partial f}{\partial r_s}\frac{\partial r_s}{\partial t}
+ v_s^2 f \frac{\partial f}{\partial r_s}\frac{\partial r_s}{\partial x}
= h(t).
\label{formeq1}
\end{equation}
The partial derivative of Eq.\ (\ref{regfunction}) may be written as
follows,
\begin{equation}
\frac{\partial f}{\partial r_s} = \frac{\sigma}{2 \tanh{(\sigma R)}} 
\left\{\sech^2[\sigma(r_s + R)] - \sech^2[\sigma(r_s - R)]\right\}.
\end{equation}
The partial derivatives of $r_s$ yield,
\begin{align}
\frac{\partial r_s}{\partial t} &= \pm \frac{d x_s}{dt} = \pm v_s(t),
\\
\frac{\partial r_s}{\partial x} &= \pm 1,
\end{align}
and remembering that in the dust case $\beta = \beta(x,t)$, then
$r_s(t)$ is given by (see Eq.\ \ref{rsxyz}),
\begin{equation}
r_s(t) = \sqrt{\left[x - x_s(t)\right]^2} = 
\lvert x - x_s(t) \rvert.
\end{equation}
Considering the expressions above, Eq.\ \eqref{formeq1} may be
rewritten as follows,
\begin{equation}
f \frac{d^2 x_s}{dt^2}
\pm v_s^2 \frac{\sigma F}{2 \tanh{(\sigma R)}}
\pm v_s^2 f \frac{\sigma F}{2 \tanh{(\sigma R)}} = h(t),
\label{formeq2}
\end{equation}
where
\begin{equation} 
F(r_s) \equiv \sech^2[\sigma(r_s + R)]-\sech^2[\sigma(r_s - R)].
\end{equation}
Remembering that the regulating function $f[r_s(t)]$, as defined by Eq.\
\eqref{regfunction}, can be approximated by a top hat function when
$\sigma \gg R$ (see Ref.\ \cite{nos}, \S 2.1), in this limit Eq.\
\eqref{formeq2} takes the following form inside the bubble,
\begin{equation}
\frac{d^2 x_s}{dt^2} = h(t),
\end{equation}
where $f=1$. Outside the bubble $f = 0$ and then $h(t) = 0$, which means
plane shock waves described by the inviscid Burgers equation \cite{nos}.

Hence, the nonuniqueness of the shift vector $\beta = v_s(r_s) f[r_s(t)]$
arises from the fact that not only the function $h(t)$ is arbitrary, but
also because the regulating function $f[r_s(t)]$ only requires a top hat
behavior with null values outside the bubble. So, any well behaved
function that respects such constraints may be part of a solution of the
Burgers equation. Moreover, from the energy conditions calculated for the
PPF, as summarized in Table \ref{tab3}, one can see that $\beta$ plays a
fundamental role in Cases 1a and 2a for the null and dominant energy
conditions to be satisfied, which adds further physical constraints to
its behavior. So, it is clear that the nonlinearity of the Einstein
equations imply that the generic functions appearing in their
integration become entangled with the regulating function in a non
trivial manner.

To extend this analysis to the perfect fluid and PPF contents require
further constraints on the shift vector, something which at this stage
would be done in an entirely arbitrary manner. Perhaps in the future
that can be done under more physically plausible reasoning.

\subsection{Anisotropic Fluids}\lb{anifluid}

The PPF proposed in Section \ref{diffluid} aimed at offering an
alternative EMT for solving the Einstein equations endowed with the
warp drive metric. As we shall see below, The PPF can actually be seen
as an anisotropic fluid with heat flux \cite{CarlEckart1940}.

In general relativity the energy momentum tensor $T_{\mu\nu}$ represents
the source of energy and momentum, where $T_{00}$ is the flow of energy
across a surface of constant time (energy density), $T_{0 i}$ is the
energy flux across a surface in the $i$ direction (constant $x^i$),
$T_{i 0}$ is the momentum density and $T_{ij}$ is the momentum flux. If
we choose a comoving frame of reference that moves with the same
velocity as the fluid this means that particles in this fluid will have
zero velocity and the flux of energy will be only through the flux of
heat, and the momentum flux will be via some sort of dissipative
phenomena such as viscosity, thermal radiation or even some sort of
electromagnetic type of radiation.

The general stress-energy tensor of a relativistic fluid can be written
in the form below \cite{CarlEckart1940,Pimentel2016},
\be
 T^{\alpha \beta} = \mu u^\alpha u^\beta + p h^{\alpha \beta} 
+ u^\alpha q^\beta + u^\beta q^\alpha  + \pi^{\alpha \beta},
\ee
where
\be
h_{\alpha \beta} = g_{\alpha \beta} + u_a\,u_b \,,
\ee
projects tensors onto hypersurfaces orthogonal to $u^\alpha$, $\mu$ is
the matter density, $p$ is the fluid static pressure, $q^\alpha$ is the
heat flux vector and $\pi^{\alpha \beta}$ is the viscous shear tensor.
The world lines of the fluid elements are the integral curves of the
four-velocity vector $u^\alpha$. The heat flux vector and viscous shear
tensor are transverse to the world lines, that is, 
\be
q_a\,u^a = 0 , \;\; \pi_{ab}\,u^b = 0.
\ee

In terms of coordinates we can write the energy momentum tensor for
a general fluid as
\be
T_{\alpha\beta} = 
\begin{pmatrix} 
\varepsilon & q_a  \\ 
q_b   & \pi_{ab}  
\end{pmatrix},
\ee
where $q_a$ is the three-vector heat flux vector, $\varepsilon$ is a
scalar function and $\pi_{ab}$ is a three by three matrix viscous
stress tensor, which is symmetric and traceless. Both $q_a$ and
$\pi_{ab}$ have respectively three and five linearly independent
components. Together with the density $\mu$ and the static pressure
$p$, this makes a total of ten linearly independent components which
is the number of linearly independent components in a four-dimensional
symmetric rank two tensor. We noticed that the Einstein tensor
components are highly non-linear for the warp drive metric and the
off diagonal terms require those free parameters for a non over
determined solution. For a non curved metric, i.e., the Minkowski
metric $\eta_{\alpha \beta}$, the energy momentum tensor for a perfect
fluid with anisotropic pressures can be written as 
\be
T_{\alpha\beta} = 
\begin{pmatrix} 
\mu & 0 & 0 & 0  \\ 
0   & p_x & 0 & 0  \\ 
0   & 0 & p_y & 0  \\ 
0   & 0 & 0 & p_z 
\end{pmatrix}.
\ee
Isotropic static pressure means that $p_x=p_y=p_z=p$. The perfect
fluid has no heat flux or dissipative phenomena, then $(q^\alpha=0,
\pi^{\alpha\beta}=0)$. This special case with dust content is the
well-known EMT, that is, 
\be
T^{\alpha \beta} = \mu u^\alpha u^\beta + p h^{\alpha \beta}
= (\mu + p) u^\alpha u^\beta + p g^{\alpha \beta}\,.
\ee

The PPF proposed in Section \ref{diffluid} has the matrix form given by
Eq.\ (\ref{71}), which may be broken down as the sum of a perfect fluid
in a warp drive background as given by Eq.\ (\ref{pfemt}), and a
dissipative fluid with the heat flux four-vector given by, 
$q_\alpha =-\frac{1}{2}(q_0, q_1, 0, 0)$. Hence,
\be
q_\alpha u_\beta + q_\beta u_\alpha = 
\begin{pmatrix} 
q_0 & q_1 & 0 & 0  \\ 
q_1 & 0 & 0 & 0  \\ 
0   & 0 & 0 & 0  \\ 
0   & 0 & 0 & 0 
\end{pmatrix},
\ee
since the four-velocity for the moving frame is $u_\alpha = (-1,0,0,0)$.
The isotropic term is given by,
\be
\pi_{\alpha\beta} = 
\begin{pmatrix} 
\pi_{00} & 0 & 0 & 0  \\ 
0      & \pi_{01} & 0 & 0  \\ 
0              & 0 & \pi_{02} & 0  \\ 
0              & 0 & 0 & \pi_{03} 
\end{pmatrix},
\ee
and
\be
\mu + \beta^2 p \to (\mu + \pi_{00} + q_0) + \beta^2 p  \,.
\ee
So, the four parameters of the PPF are as follows, 
\be
A = \pi_{01} + p\,,
\ee
\be
B = \pi_{02} + p\,,
\ee
\be
C = \pi_{03} + p\,,
\ee
\be
D = p - \frac{q_1}{\beta}.
\ee
The tensor $\pi_{\alpha\beta}$ must be 
traceless, giving us one more equation to solve for the
free parameters (anisotropic pressures)
\be
\pi_{00} + \pi_{01} + \pi_{02} + \pi_{03} = 0
\ee

From the above one can see that the warp drive metric endowed with
the PPF allows the study of fluid anisotropy coupled with possible
dissipative effects that could lead to a warp drive bubble. Perfect
fluids are well known to be part of solutions for the Einstein
equations, this being the case of the standard FLRW cosmological
model that accounts for the expansion, isotropy and spatial homogeneity
of the universe. On the other hand, anisotropic imperfect fluids
offer a more complex source of gravitational effects, presenting
dissipative processes, shear and bulk viscous pressures, interaction
between fluids, radiation processes, electromagnetic interaction and
even collision between particles, charged or not. These fluids are
even known to avoid the big bang singularity in cosmological models
\cite{murphy,klimek,msc,p1,p2}. MacCallum \cite{MacCallum1979}
discussed various ways of generating anisotropy such as the presence
of electromagnetic fields, the presence of viscous terms and the
anisotropic stresses due to the anisotropic expansion of a cloud of
collisionless particles. Another way to account for viscosity, heat
and energy flux is the interaction of two or more fluids
\cite{Letelier1980,Bayin1982}.

\subsection{Other aspects of warp drive physics}

The points presented above regarding the physical feasibility of
superluminal travel with the Alcubierre spacetime geometry by no
means exhaust this discussion. Several issues remain open, with each
of them deserving separate studies that are beyond the scope of this
paper, since in here we focused on the basic properties of the
Einstein equations' solutions of the warp drive metric with fluid
content. With respect to the open issues, one can point out the
amount of mass-energy density, exotic or not, necessary for the
feasibility for the warp drive in the context of both the perfect
fluid and the PPF, as well as quantum effects and the question of
stability or instability in our solutions. These issues deserve
further investigations and are the subject of ongoing research.

\section{Conclusions}\label{conc}
\renewcommand{\theequation}{8.\arabic{equation}}
\setcounter{equation}{0}

In this work we have analyzed the Einstein equations for the Alcubierre
warp drive metric having as gravity source two types of energy-momentum
tensors (EMT) for fluid, namely the perfect fluid and the parametrized
perfect fluid (PPF). The latter is defined by allowing the EMT
pressure components of the perfect fluid to be different from one another
and dependent on all coordinates.

After obtaining the components of the Einstein tensor for the warp drive
metric we calculated the dynamic equations for both fluids by solving
the respective Einstein equations. Solutions were found in the form of
various equations of state, and, by further imposing the null divergence
for the EMTs, new constraints were also found for the various variables.
The weak, dominant, strong and null energy conditions were also
calculated, which implied further constraints upon the free quantities
for the EMTs. For the perfect fluid these were on the matter density
$\mu$ and pressure $p$. For the PPF that occurred on the pressure
components $p, A, B, C$, the matter density for the fluid $\mu$ and the
momentum component $D$.

We found two main groups of solutions subcases possessing different
conditions for each EMT. For the perfect fluid, one solution  may be
interpreted as requiring that the warp bubble can only be viable with
negative matter density. The alternative interpretation is that the
warp bubble is possible with positive matter density, but in this
case the regulating function $f(r,s)$, which shapes the warp bubble,
becomes a complex function. This comes from results allowing the
possibility that the function $\beta=v_s(t) f(r_s)$ may have complex
solutions once the matter density is positive. Other results in
both fluids reinforce our earlier finding in Ref.\ \cite{nos} that
the warp bubble necessary for generating superluminal velocities, or
warp speeds, can be interpreted as a shock wave from classical fluid
dynamics theory.

We named four cases arising from the solutions of the Einstein
equations: 1a, 1b, 2a and 2b. However, the Cases 1a and 2a are very
similar, or equal, to each other, the same happening to Cases 1b and
2b. For this reason they were grouped together in the tables that
summarize all results.

Specifically, Cases 1b and 2b for the perfect fluid reduced the
solutions to the ones found for dust content already studied in Ref.\
\cite{nos}, that is, a vacuum solution unable to create a warp bubble,
but which connects the warp metric to the inviscid Burgers equation,
also yielding $\beta$ as a function of $t$ and $x$ coordinates. The
null EMT divergence is satisfied and a continuity equation was also
found (Eq.\ \ref{divpf0}).

Cases 1b and 2b for the PPF resulted in a equation of state of the
form $\mu =-\beta^2 p$, coordinate dependency of the $\beta$ function
became $\beta=\beta(x,t)$ and a non-homogeneous Burgers equation
(\ref{c2bfinal03}) emerged. However, the pressures were
constrained in such a way that this PPF EMT solution for the warp
drive was dismissed as unphysical. 

Cases 1a and 2a for the perfect fluid produced an equation of state
relating pressure and matter density given by $p=3\mu$. The $\beta$
function dependencies became $\beta=\beta(y,t)$ in the former case
and $\beta=\beta(z,t)$ in the latter, and both produced a differential
equation for $\beta$ that either requires negative density for
$\beta$ to be a real valued function, or a positive matter density
which then leads to a complex solution for $\beta$, which in turn
leads to a complex regulating function $f(r_s)$ whose possible real
part would then be related to a physically viable warp bubble.

Cases 1a and 2a for the PPF resulted in an equation of state relating
almost all quantities, in the form $\mu=\beta^2(2D-A-p)+A/3$ which
is valid for both cases. Coordinate dependency on $\beta$ became,
respectively, resulted in $\beta=\beta(y,t)$ and $\beta=\beta(z,t)$.
The function $\beta$ is also governed by the first order differential
equations (\ref{abc1}) and (\ref{abc2}), respectively.

The null EMT divergence $T{^{\alpha\sigma}}_{;\sigma}=0$ were
calculated, producing further sets of very nonlinear differential
equations constraining all quantities in the PPF which could be used,
in principle, to determine all pressures and matter density in this
fluid. For the perfect fluid, Cases 1a and 2a are reduced to a
continuity equation (\ref{divpf0}) including the function $\beta$,
which is interpreted as playing the role of the flow velocity vector
field. Cases 1b and 2b reduced the PPF to either the trivial condition
of Minkowski flat spacetime with no warp drive, or an EMT with all
components being zero, that is a vacuum case. 

It has already been seen in Ref.\ \cite{nos} that the Burgers equation 
appears connected to the dust solution of the warp drive metric,
which is in fact a vacuum solution. Cases 1b and 2b of the perfect
fluid became reduced to the dust solution, and a non-homogeneous
form of the Burgers equation appears in these respective cases for
the PPF, although the whole solutions in these cases were dismissed
as unphysical. The solutions that presented themselves as the most
plausible ones for creating warp speeds, 1a and 2a for both the
perfect fluid and PPF, do not generate a Burgers equation.

The weak, dominant, strong and null energy conditions were also studied
in the context of the perfect fluid and PPF energy-momentum tensors for
a warp drive metric. The resulting expressions were found to satisfy
all conditions in the perfect fluid EMT. Regarding the PPF, specific 
expressions constraining its EMT quantities were obtained in order to
satisfy these energy conditions, but they do not necessarily lead to
the conclusion that negative matter density is always necessary for
viable warp speeds, particularly because in the PPF the pressure $A$
must assume values equal to zero or positive.

Summing up, the results of this paper indicate that warp speeds
might be physically viable in the context of positive matter density
as some solutions of the Einstein equations for both fluids keep
open this possibility.  Nevertheless, such a situation creates the
additional issue in the perfect fluid context concerning the meaning
of a possible complex regulating function in the warp metric, a
result that may be interpreted as a caveat, or major stumbling
block. Such difficulty does not appear to happen in the PPF
scenario, although this fluid was considered here mainly as a
hypothetical model whose aim was to investigate whether or not
new possibilities arise in the solutions of the Einstein field
equations considering more complex energy-momentum tensors having
the Alcubierre warp drive metric. On this front it seems then that
the initial conclusions about the unphysical nature of warp drive,
or the impossibility of generating warp speeds, may not be not as
stringent as initially thought, or, perhaps, not valid at all.

\section*{Acknowledgments}

\ni We are grateful to the referees for useful comments. E.M.C.A.\
thanks CNPq (Conselho Nacional de Desenvolvimento Cient\'ifico e
Tecnol\'ogico), Brazilian scientific support federal agency, for
partial financial support, grants numbers 406894/2018-3 and 302155/2015-5. 


\end{document}